\documentclass[preprintnumbers]{revtex4-1}
\usepackage{graphicx}

\newcommand{\beq}{\begin{equation}}
\newcommand{\eeq}{\end{equation}}
\newcommand{\diag}{{\rm diag}}
\newcommand{\trace}{{\rm trace}}
\newcommand{\bi}{\begin{itemize}}
\newcommand{\ei}{\end{itemize}}
\newcommand{\ba}{\begin{array}}
\newcommand{\ea}{\end{array}}
\newcommand{\Hz}{H\!z}
\newcommand{\Ploss}{\Pi_{\mathrm{loss}}}

\usepackage{multirow}
\usepackage{subfig}  %Use as needed
\usepackage{algorithm}
\usepackage{algorithmic}
\usepackage{tikz}
\usepackage{pgf,pgfplots}
\usetikzlibrary{plotmarks}
\pgfplotsset{compat=newest}
\pgfplotsset{
	ylabel style={yshift=-0.25em}
		  }

\usepackage{amssymb,amsmath}

\usepackage{url}

\newcommand{\cL}{{\cal L}}
\newcommand{\aL}{{\cal L}_\rho}

\newcommand{\bbR}{\mathbb{R}}
\newcommand{\bbC}{\mathbb{C}}

\newcommand{\non}{\nonumber}
\newcommand{\ds}{\displaystyle}

\newcommand{\Fdmd}{F_{\mathrm{dmd}}}
\newcommand{\Vand}{V_{\mathrm{and}}}

\newcommand{\Dalpha}{D_{\alpha}}
\newcommand{\Dbeta}{D_{\beta}}
\newcommand{\DefinedAs}[0]{\mathrel{\mathop:}=}
\begin{document}

    \preprint{Submitted to Physics of Fluids}

\title{Sparsity-promoting dynamic mode decomposition}

\author{Mihailo R.\ Jovanovi\'c}
    \homepage{http://umn.edu/home/mihailo/}
    \email{mihailo@umn.edu}
    \affiliation{%
    Department of Electrical and Computer Engineering, University of
    Minnesota,~Minneapolis,~MN~55455,~USA%
    }%
\author{Peter J.\ Schmid}
    \homepage{http://yakari.polytechnique.fr/people/peter/}
    \email{peter.schmid@ladhyx.polytechnique.fr}
    \affiliation{%
    Laboratoire d'Hydrodynamique (LadHyX),~Ecole Polytechnique,~91128 Palaiseau cedex,~France %
    }%
\author{Joseph W.\ Nichols}
    \homepage{http://www.aem.umn.edu/people/faculty/bio/nichols.shtml}
    \email{jwn@umn.edu}
    \affiliation{%
    Department of Aerospace Engineering and Mechanics, University of
    Minnesota,~Minneapolis,~MN~55455,~USA%
    }%
 
\date{\today}% It is always \today, today,

	\begin{abstract}
Dynamic mode decomposition (DMD) represents an effective means for capturing the essential features of numerically or experimentally generated flow fields. In order to achieve a desirable tradeoff between the quality of approximation and the number of modes that are used to approximate the given fields, we develop a sparsity-promoting variant of the standard DMD algorithm. In our method, sparsity is induced by regularizing the least-squares deviation between the matrix of snapshots and the linear combination of DMD modes with an additional term that penalizes the $\ell_1$-norm of the vector of DMD amplitudes. The globally optimal solution of the resulting regularized convex optimization problem is computed using the alternating direction method of multipliers, an algorithm well-suited for large problems. Several examples of flow fields resulting from numerical simulations and physical experiments are used to illustrate the effectiveness of the developed method.
	\end{abstract}

\maketitle

\newpage

\section{Introduction}
	\vspace*{-2ex}
	\label{sec.intro}
	
Even though fluid flows are infinite-dimensional systems governed by
nonlinear partial differential equations, the essential features of
their dynamical responses can often be approximated reliably by models
of low complexity. This observation has given rise to the notion of
coherent structures -- organized fluid elements that, together with
dynamic processes, are responsible for the bulk of momentum and energy
transfer in the flow. Recent decades have witnessed significant
advances in the extraction of coherent structures from data collected
in experiments and numerical simulations. For example, proper
orthogonal decomposition (POD) modes~\cite{lumley,sirovich}, global
eigenmodes, frequential modes~\cite{sipp}, and balanced
modes~\cite{moore,rowley} have provided useful insight into the
dynamics of fluid flows. Recently, Koopman
modes~\cite{Mezic05,rowley2,Mezic13} and dynamic mode decomposition
(DMD)~\cite{schJFM10dmd} have joined the group of feature extraction
techniques. Both POD and DMD are snapshot-based post-processing
algorithms which may be applied equally well to data obtained in
simulations or in experiments. While POD modes are characterized by
spatial orthogonality and multi-frequential temporal content, DMD
modes may be non-orthogonal but each of them possesses a single temporal
frequency. This lack of non-orthogonality of DMD modes may be
essential to capturing important dynamical effects in systems with
non-normal dynamical generators~\cite{tretrereddri93,jovbamJFM05,Schmid2007}. For an in-depth discussion of the connection between DMD and other data decomposition methods, we refer the reader to~\cite{schJFM10dmd,Mezic13,bag13}.

The importance of POD and DMD modes goes beyond identification of coherent structures in fluids flows. In particular, they may be used to obtain models of low complexity; by projecting the full system onto the subspace spanned by the extracted modes, the governing equations may be approximated by a dynamical system with fewer number of degrees of freedom. This facilitates computationally tractable study of flow stability and receptivity as well as a model-based control design. In many situations, however, it is challenging to identify a subset of modes that have the strongest impact on the flow dynamics. For example, spatial non-orthogonality of the DMD modes may introduce a poor quality of approximation of experimentally or numerically generated snapshots when only a subset of modes with the largest amplitude is retained. Recent attempts at extracting only a subset of desired frequencies and spatial profiles rely on formulation of non-convex optimization problems. In~\cite{cheturow12}, a variant of DMD, referred to as the Optimized DMD, was introduced; determining a solution to this problem in general requires an intractable combinatorial search. In~\cite{gouwynpeaCDC12,wynpeagangouJFM13}, a gradient-based algorithm was employed to simultaneously search for the low-rank basis and the matrix that governs temporal evolution on a lower-dimensional subspace in order to reduce the least-squares residual achieved by DMD.

In this paper, we develop a sparsity-promoting variant of the standard DMD algorithm. This algorithm is aimed at achieving a desirable tradeoff between the quality of approximation (in the least-squares sense) and the number of modes that are used to approximate numerical or experimental snapshot sequences. To achieve this objective, we combine tools and ideas from convex optimization~\cite{boyd} with the emerging area of compressive sensing~\cite{canromtao06,don06,cantao06}. Our approach to inducing sparsity relies on regularization of the least-squares deviation (between the matrix of snapshots and the linear combination of DMD modes) with an additional term that penalizes the $\ell_1$-norm of the vector of DMD amplitudes. The $\ell_1$-norm can be interpreted as a convex relaxation of the non-convex cardinality function~\cite{boyd}, and it has been effectively used as a proxy for promoting sparsity in a number of applications~\cite{cantao06,canwakboy08,hastibfri09,boyparchupeleck11,linfarjovTAC13admm}. The alternating direction method of multipliers (ADMM) -- a state-of-the-art algorithm for solving large-scale and distributed optimization problems~\cite{boyparchupeleck11} -- is then employed to solve the resulting convex optimization problem and to efficiently compute the globally optimal solution. Since this algorithm alternates between promoting sparsity and minimizing the least-squares residual, we exploit the respective structures of the underlying penalty functions to decompose the optimization problem into easily solvable modules. In particular, we show that the least-squares minimization step amounts to solving an unconstrained regularized quadratic program and that sparsity is promoted through the application of a convenient soft-thresholding operator. After a desirable tradeoff between the quality of approximation and the number of DMD modes has been achieved, we fix the sparsity structure and compute the optimal amplitudes of the retained dynamic modes as the solution to the constrained quadratic program.

Our presentation is organized as follows. In Section~\ref{sec.problem}, we formulate the problem, provide a brief overview of the dynamic mode decomposition, and address the optimal selection of amplitudes of extracted DMD modes. In Section~\ref{sec.sparse}, we apply a sparsity-promoting framework to select a subset of DMD modes which strikes a user-defined balance between the approximation error (with respect to the full data sequence) and the number of extracted modes. In Section~\ref{sec.examples}, we use three databases resulting from the two-dimensional linearized Navier--Stokes equations for plane Poiseuille flow, the unstructured large-eddy simulation (LES) of a supersonic jet, and the time-resolved particle image velocimetry (TR-PIV) experiment of a flow through a cylinder bundle to illustrate the utility of the developed method. We conclude our presentation in Section~\ref{sec.remarks} with a summary of our contributions and an outlook for future research directions, and relegate algorithmic developments to the Appendices.

	\vspace*{-2ex}
\section{Problem formulation}
	\vspace*{-1ex}
    \label{sec.problem}

\subsection{Dynamic mode decomposition}
	\vspace*{-1ex}
	
The dynamic mode decomposition (DMD) is a data processing algorithm
that extracts coherent structures with a single frequency from a
numerical or experimental data-sequence~\cite{schJFM10dmd}. In what
follows, we briefly outline the key steps of DMD.

We begin by collecting a sequence of snapshots from numerical
simulations or physical experiments and form a data matrix whose
columns represent the individual data samples. Even though we confine
our attention to temporal evolution processes, the DMD-framework can
accommodate a variety of ``evolution coordinates'' (e.g., spatial
directions, or curved base-flow
streamlines)~\cite{schJFM10dmd}. Furthermore, we assume that the data are
equispaced in time, with a time step $\Delta t$,
\[
\left\{
  \psi_0, \psi_1, \ldots, \psi_N
\right\},
\]
where each $\psi_i \DefinedAs \psi (i \Delta t) $ is, in general, a
complex vector with $M$ components (measurement points), i.e., $\psi_i
\in {\mathbb{C}}^M$.

Next, we form two data matrices from the snapshot sequence
\[
\begin{array}{rcl}
\Psi_0
& \DefinedAs &
\left[
  \begin{array}{cccc}
  \psi_0
  &
  \psi_1
  &
  \cdots
  &
  \psi_{N-1}
  \end{array}
\right]
\, \in \; {\mathbb{C}}^{M \times N},
\\[0.25cm]
\Psi_1
& \DefinedAs &
\left[
  \begin{array}{cccc}
  \psi_1
  &
  \psi_2
  &
  \cdots
  &
  \psi_{N}
  \end{array}
\right]
\, \in \; {\mathbb{C}}^{M \times N},
\end{array}
\]
and postulate that the snapshots have been generated by a
discrete-time linear time-invariant system
\begin{equation}
\psi_{t + 1}
\; = \;
A \, \psi_t,
~~
t
\, = \,
\left\{
  0, \ldots, N - 1
\right\}.
\label{eq.lti}
\end{equation}
For fluid flows, the matrix $A$ typically contains a large number of
entries (which are in general complex numbers). The dynamic mode
decomposition furnishes a procedure for determining a low-order
representation of the matrix $A \in {\mathbb{C}}^{M \times M}$ that
captures the dynamics inherent in the data sequence. In fluid
problems, the number of components (measurement points) in each
snapshot $\psi_i$ is typically much larger than the number of
snapshots, $M \gg N,$ thereby implying that $\Psi_0$ and $\Psi_1$ are
tall rectangular matrices. Using the linear relation~(\ref{eq.lti})
between the snapshots at two consecutive time steps, we can link the
two data matrices $\Psi_0$ and $\Psi_1$ via the matrix $A$ and express
$\Psi_1$ as
\begin{equation}
    \begin{array}{rcl}
    \Psi_1
    & = &
    \left[
    \begin{array}{cccc}
    \psi_1
    &
    \psi_2
    &
    \cdots
    &
    \psi_{N}
    \end{array}
    \right]
    \\[0.25cm]
    & = &
    \left[
    \begin{array}{cccc}
    A \, \psi_0
    &
    A \, \psi_1
    &
    \cdots
    &
    A \, \psi_{N-1}
    \end{array}
    \right]
    \\[0.25cm]
    & = &
    A \, \Psi_0.
    \end{array}
    \label{eq.PsiPsi}
\end{equation}

For a rank-$r$ matrix of snapshots $\Psi_0$, the DMD algorithm
provides an optimal representation $F \in \bbC^{r \times r}$ of the
matrix $A$ in the basis spanned by the POD modes of $\Psi_0$,
\begin{equation}
  A
  \; = \;
  U \, F \, U^*.
  \non
\end{equation}
Here, $U^*$ denotes the complex-conjugate-transpose of the matrix of
POD modes $U$ which is obtained from an economy-size singular value
decomposition (SVD) of $\Psi_0 \in {\mathbb{C}}^{M \times N}$,
\begin{equation}
  \Psi_0
  \; = \;
  U \, \Sigma \, V^*,
  \non
\end{equation}
where $\Sigma$ is an $r \times r$ diagonal matrix with non-zero
singular values $\{ \sigma_1, \ldots, \sigma_r \}$ on its main
diagonal, and
\[
    \begin{array}{rcl}
    U \, \in \, {\mathbb{C}}^{M \times r}
    &
    \mbox{with}
    &
    U^* \, U
    \; = \;
    I,
    \\[0.25cm]
    V \, \in \, {\mathbb{C}}^{r \times N}
    &
    \mbox{with}
    &
    V^* \, V
    \; = \;
    I.
    \end{array}
\]
The matrix $F$ can be determined from the matrices of snapshots
$\Psi_0$ and $\Psi_1$ by minimizing the Frobenius norm of the
difference between $\Psi_1$ and $A \, \Psi_0$ with $ A \, = \, U \, F
\, U^*$ and $\Psi_0 \, = \, U \, \Sigma \, V^*$,
\begin{equation}
  \underset{F}{\text{minimize}}
  ~~
  \|
  \Psi_1
  \; - \;
  U \, F \, \Sigma \, V^*
  \|_F^2,
  \label{eq.optF}
\end{equation}
where the Frobenius norm of the matrix $Q$ is determined by
\[
    \|
    Q
    \|_F^2
    \; = \;
    \hbox{trace}
    \left(
    Q^* \, Q
    \right)
    \, = \;
    \hbox{trace}
    \left(
    Q \, Q^*
    \right).
\]
It is straightforward to show that the optimal solution
to~(\ref{eq.optF}) is determined by
\begin{equation}
    \Fdmd
    \; = \;
    U^* \, \Psi_1 \, V \, \Sigma^{-1}.
    \non
\end{equation}
This expression is identical to the expression provided
in~\cite{schJFM10dmd} and it concludes the implementation of the DMD
algorithm, starting from matrices of data snapshots $\Psi_0$ and
$\Psi_1$. For a discussion about the relation between $\Fdmd$ and the
companion form matrix $A_c$ -- which provides a representation of the
matrix $A$ on an $N$-dimensional subspace of ${\mathbb{C}}^M$ that is
spanned by the columns of $\Psi_0$ -- we refer the reader
to~\cite{schJFM10dmd}.

	\vspace*{-2ex}
\subsection{Optimal amplitudes of DMD modes}
	\vspace*{-1ex}
\label{sec.amplitudes}

The matrix $\Fdmd \in \bbC^{r \times r}$ determines an optimal low-dimensional
representation of the inter-snapshot mapping $A \in \bbC^{M \times M}$ on the subspace spanned by the POD modes of $\Psi_0$. The
  dynamics on this $r$-dimensional subspace are governed by
  \begin{equation}
    x_{t + 1}
    \; = \;
    \Fdmd \, x_t,
    \label{eq.x}
  \end{equation}
  and the matrix of POD modes $U$ can be used to map $x_t$ into a
  higher dimensional space $\bbC^M$,
  \[
  \psi_t \; \approx \; U \, x_t.
  \]
  If $\Fdmd$ has a full set of linearly independent eigenvectors $\{
  y_1, \ldots, y_r \}$, with corresponding eigenvalues $\{ \mu_1,
  \ldots, \mu_r \}$, then it can be brought into a diagonal coordinate
  form,
  \begin{equation}
    \Fdmd
    \; = \;
    \underbrace{
      \left[
    \begin{array}{ccc}
    y_1 & \cdots & y_r
    \end{array}
    \right]}_Y
    \underbrace{\left[ \begin{array}{ccc} \mu_1 & & \\ & \ddots & \\ & & \mu_r \end{array} \right]}_{D_{\mu}}
     \underbrace{\left[\begin{array}{c} ~~z_1^*~~ \\ ~~\vdots~~ \\ ~~z_r^*~~ \end{array}\right]}_{Z^*}.
    \non
  \end{equation}
  Here, $\{ z_1, \ldots, z_r \}$ are the eigenvectors of $\Fdmd^*$,
  corresponding to the eigenvalues $\{ \bar{\mu}_1, \ldots,
  \bar{\mu}_r \}$, which are suitably scaled so that the following
  bi-orthogonality condition holds
  \[
  z_i^* \, y_j \; = \, \left\{ \ba{rl} 1, & i \, = \, j
    \\[0.15cm]
    0, & i \, \neq \, j.  \ea \right.
  \]
  Now, the solution to~(\ref{eq.x}) is determined by
  \[
  x_t \; = \; Y \, D_{\mu}^t \, Z^* x_0 \; = \; \sum_{i \, = \, 1}^{r}
  \, y_i \, \mu_i^t \, z_i^* x_0 \; = \; \sum_{i \, = \, 1}^{r} \, y_i
  \, \mu_i^t \, \alpha_i,
  \]
where $\alpha_i \DefinedAs z_i^* x_0$ represents the $i$th modal
  contribution of the initial condition $x_0$. We can thus approximate
  experimental or numerical snapshots using a linear combination of
  the DMD modes, $\phi_i \DefinedAs U y_i$,
  \begin{equation}
    \psi_t
    \; \approx \;
    U \, x_t
    \; = \;
    \sum_{i \, = \, 1}^{r}
    \,
    \phi_i
    \,
    \mu_i^t
    \,
    \alpha_i,
    ~~
    t
    \; \in \,
    \left\{
      0, \ldots, N - 1
    \right\},
    \non
  \end{equation}
  and each $\alpha_i$ can be interpreted as the ``amplitude'' of the
  corresponding DMD mode~\cite{schJFM10dmd}. Equivalently, in matrix form,
  we have
\begin{equation}
  \underbrace{
    \left[
      \begin{array}{cccc}
        \psi_0
        &
        \psi_1
        &
        \cdots
        &
        \psi_{N-1}
      \end{array}
    \right]
  }_{\Psi_0}
  ~ \approx ~
  \underbrace{
    \left[
      \begin{array}{cccc}
        \phi_1
        &
        \phi_2
        &
        \cdots
        &
        \phi_{r}
      \end{array}
    \right]
  }_{\Phi}
  \,
  \underbrace{
    \left[
      \begin{array}{cccc}
        \alpha_1
        &
        &
        &
        \\
        &
        \alpha_2
        &
        &
        \\
        &
        &
        \ddots
        &
        \\
        &
        &
        &
        \alpha_r
      \end{array}
    \right]
  }_{D_{\alpha} \; \DefinedAs ~ \hbox{diag} \, \{ \alpha \}}
  \,
  \underbrace{
    \left[
      \begin{array}{cccc}
        1
        &
        \mu_1
        &
        \cdots
        &
        \mu_1^{N-1}
        \\[0.15cm]
        1
        &
        \mu_2
        &
        \cdots
        &
        \mu_2^{N-1}
        \\
        \vdots
        &
        \vdots
        &
        \ddots
        &
        \vdots
        \\[0.15cm]
        1
        &
        \mu_r
        &
        \cdots
        &
        \mu_r^{N-1}
      \end{array}
    \right]
  }_{\Vand},
  \non
\end{equation}
which demonstrates that the temporal evolution of the dynamic modes is
governed by the Vandermonde matrix $\Vand \in \bbC^{r \times N}$. This
matrix is determined by the $r$ complex eigenvalues $\mu_i$ of $\Fdmd$
which contain information about the underlying temporal frequencies
and growth/decay rates.

Determination of the unknown vector of amplitudes $\alpha \DefinedAs
\left[ \begin{array}{ccc} \alpha_1 & \cdots & \alpha_r \end{array}
\right]^T$ then amounts to finding the solution to the following
optimization problem
\begin{equation}
  \begin{array}{rl}
    \underset{\alpha}{\text{minimize}}
    &
    \|
    \Psi_0
    \; - \;
    \Phi \, \Dalpha \, \Vand
    \|_F^2.
  \end{array}
  \non
\end{equation}
Using the economy-size SVD of $\Psi_0 = U \, \Sigma \, V^*$ and the
definition of the matrix $\Phi \DefinedAs U \, Y$, we bring this
problem into the following form
\begin{equation}
  \begin{array}{rl}
    \underset{\alpha}{\text{minimize}}
    &
    J (\alpha)
    \; \DefinedAs \;
    \|
    \Sigma \, V^*
    \; - \;
    Y \, \Dalpha \, \Vand
    \|_F^2,
  \end{array}
  \label{eq.alpha_opt}
\end{equation}
which is a convex optimization problem that can be solved using
standard methods~\cite{boyd,cvx}. We note that this
  optimization problem does not require access to the POD modes of the
  matrix of snapshots $\Psi_0$; the problem data
  in~(\ref{eq.alpha_opt}) are the matrices $\Sigma$ and $V$, which are
  obtained from the economy-size SVD of $\Psi_0$, and the matrices $Y$
  and $\Vand$, which result from the eigenvalue decomposition of
  $\Fdmd$.

In Appendix~\ref{sec.ls-amplitudes}, we show that the objective
function $J (\alpha)$ in~(\ref{eq.alpha_opt}) can be equivalently
represented as
	\beq 
	J (\alpha) 
	\; = \; 
	\alpha^* P \, \alpha 
	\; - \; 
	q^* \alpha 
	\; - \; 
	\alpha^* q 
	\; + \; 
	s,
\label{eq.Jvec}
	\eeq 
where
	\beq 
	P 
	\; \DefinedAs \, 
	\left( Y^* \, Y \right) 
	\circ 
	\left( \overline{\Vand \, \Vand^*} \right), 
  	~~ 
	q 
	\; \DefinedAs \,
	\overline{\diag \left( \Vand \, V \, \Sigma^* \, Y \right)}, 
	~~ 
	s 
	\; \DefinedAs \,
	\trace \left( \Sigma^* \Sigma \right).  
	\non
	\eeq
Here, an asterisk denotes the complex-conjugate-transpose of a vector
(matrix), an overline signifies the complex-conjugate of a vector
(matrix), $\diag$ of a vector is a diagonal matrix with its main
diagonal determined by the elements of a given vector, $\diag$ of a
matrix is a vector determined by the main diagonal of a given matrix,
and $\circ$ is the elementwise multiplication of two matrices. The
optimal vector of DMD amplitudes that solves the optimization
problem~(\ref{eq.alpha_opt}) can thus be obtained by minimizing the
quadratic function~(\ref{eq.Jvec}) with respect to $\alpha$,
\beq
\alpha_{\mathrm{dmd}} \; = \; P^{-1} q \; = \, \left( \left( Y^* \, Y
  \right) \circ \left( \overline{\Vand \, \Vand^*} \right)
\right)^{-1} \, \overline{\diag \left( \Vand \, V \, \Sigma^* \, Y
  \right)}.  \non
\eeq

A superposition of all DMD modes, properly weighted by their amplitudes and advanced in time according to to their temporal growth/decay rate, optimally approximates the entire data sequence. The key challenge that this paper addresses is the identification of a low-dimensional representation in order to capture the most important dynamic structures (by eliminating features that contribute weakly to the data sequence).

	\vspace*{-2ex}
\section{Sparsity-promoting dynamic mode decomposition}
	\vspace*{-1ex}
\label{sec.sparse}

In this section, we direct our attention to the problem of selecting
the subset of DMD modes that has the most profound influence on the
quality of approximation of a given sequence of snapshots. In other
words, we are interested in a hierarchical description of the data
sequence in terms of a set of dynamic modes. Our approach consists of
two steps. In the first step, we seek a {\em sparsity structure\/}
that achieves a user-defined tradeoff between the number of extracted
modes and the approximation error (with respect to the full data
sequence); see~(\ref{eq.Jell1}) below. In the second step, we fix the
sparsity structure of the vector of amplitudes (identified in the
first step) and determine the optimal values of the non-zero
amplitudes; see~(\ref{eq.alpha_opt_sp}) below.

We approach the problem of inducing sparsity by augmenting the
objective function $J (\alpha)$ in~(\ref{eq.alpha_opt}) with an
additional term, $\hbox{{\bf{card}}} \left( \alpha \right)$, that
penalizes the number of non-zero elements in the vector of unknown
amplitudes $\alpha$,
\begin{equation}
  \begin{array}{rl}
    \underset{\alpha}{\text{minimize}}
    &
    J (\alpha)
    \; + \;
    \gamma \, \hbox{{\bf{card}}} \left( \alpha \right).
  \end{array}
  \label{eq.Jcard}
\end{equation}
In the modified optimization problem~(\ref{eq.Jcard}), $\gamma$ is a
positive regularization parameter that reflects our emphasis on sparsity of the
vector $\alpha \in \bbC^r$. Larger values of $\gamma$ place stronger
emphasis on the number of non-zero elements in the vector $\alpha$
(relative to the quality of the least-squares approximation, $J
(\alpha)$), thereby encouraging sparser solutions to~(\ref{eq.Jcard}).

In general, finding a solution to the problem~(\ref{eq.Jcard}) amounts
to a combinatorial search that quickly becomes intractable for any
problem of interest. To circumvent this issue we introduce a relaxed
version of~(\ref{eq.Jcard}) by replacing the cardinality function with
the $\ell_1$-norm of the vector $\alpha$,
\begin{equation}
  \begin{array}{rl}
    \underset{\alpha}{\text{minimize}}
    &
    J (\alpha)
    \; + \;
    \gamma
    \,
    \ds{\sum_{i \, = \, 1}^{r}} \, | \alpha_i |.
  \end{array}
  \label{eq.Jell1}
\end{equation}
The sparsity-promoting DMD problem~(\ref{eq.Jell1}) is a {\em convex
  optimization problem\/} whose global solution, for small and medium
sizes, can be obtained using standard optimization
solvers~\cite{boyd,cvx}. In Section~\ref{sec.admm-body}, we develop an
efficient algorithm for solving~(\ref{eq.Jell1}). This algorithm
utilizes the alternating direction method of multipliers (ADMM), a
state-of-the-art method for solving large-scale and distributed
optimization problems~\cite{boyparchupeleck11}.

After a desired balance between the quality of approximation of
experimental or numerical snapshots and the number of DMD modes is
achieved, we fix the sparsity structure of the unknown vector of
amplitudes and determine only the non-zero amplitudes as the solution
to the following constrained convex optimization problem:
\begin{equation}
  \begin{array}{rl}
    \underset{\alpha}{\text{minimize}}
    &
    J (\alpha)
    \\[0.25cm]
    \text{subject to}
    &
    E^T \, \alpha
    \, = \,
    0.
  \end{array}
  \label{eq.alpha_opt_sp}
\end{equation}
In this expression, the matrix $E \in {\mathbb{R}}^{r \times m}$
encodes information about the sparsity structure of the vector
$\alpha$. The columns of $E$ are the unit vectors in $\bbR^{r}$ whose
non-zero elements correspond to zero components of $\alpha$. For
example, for $\alpha \in {\mathbb{C}}^4$ with
\[
\alpha \; = \, \left[
    \begin{array}{cccc}
    \alpha_1
    &
    0
    &
    \alpha_3
    &
    0
    \end{array}
    \right]^T,
\]
the matrix $E$ is given as
\[
E \; = \, \left[
    \begin{array}{cc}
    0 & 0
    \\
    1 & 0
    \\
    0 & 0
    \\
    0 & 1
    \end{array}
    \right].
\]

An efficient algorithm for solving~(\ref{eq.alpha_opt_sp}) is provided
in Appendix~\ref{sec.ls_sparse}.

	\vspace*{-2ex}
\subsection{Alternating direction method of multipliers}
	\vspace*{-1ex}
\label{sec.admm-body}

We next use the alternating direction method of multipliers algorithm
to find the globally optimal solution to the sparsity-promoting
optimization problem~(\ref{eq.Jell1}),
\beq
\ba{rl}
\underset{\alpha}{\text{minimize}} & J (\alpha) \; + \; \gamma \, g
(\alpha), \ea
    \label{eq.Jell1-vec}
\eeq
with $J (\alpha)$ determined by~(\ref{eq.Jvec}) and
\[
g (\alpha) \; \DefinedAs \; \ds{\sum_{i \, = \, 1}^{r}} \, | \alpha_i
|.
\]
In order to bring the problem into the form that is convenient for the
application of ADMM, we need the following two steps:
\bi
\item {\bf Step~1:} Replace the vector of amplitudes $\alpha$ in the
  sparsity-promoting term $g$ with a new variable $\beta \in \bbC^r$,
  \beq
\ba{rl} \mbox{minimize} & J (\alpha) \; + \; \gamma \, g(\beta)
  \\[0.15cm]
  \mbox{subject to} & \alpha \; - \; \beta \; = \; 0.  \ea
    \label{eq.Step1}
\eeq
For any feasible $\alpha$ and $\beta$, the optimization
problems~(\ref{eq.Jell1-vec}) and~(\ref{eq.Step1}) are
equivalent. Even though the number of optimization variables
in~(\ref{eq.Step1}) is twice as big as in~(\ref{eq.Jell1-vec}),
formulation~(\ref{eq.Step1}) allows us to exploit the respective
structures of the quadratic function $J (\alpha)$ and the
sparsity-promoting function $g (\beta)$ in the ADMM algorithm, as
outlined below.

\item {\bf Step~2:} Introduce the augmented Lagrangian,
    \beq
    \aL
    \left( \alpha,\beta,\lambda \right)
    \; \DefinedAs \;
    J(\alpha)
    \; + \;
    \gamma
    \,
    g(\beta)
    \; + \;
    \dfrac{1}{2}
    \left(
    \lambda^*
    \left(
    \alpha
    \, - \,
    \beta
    \right)
    \; + \;
    \left(
    \alpha
    \, - \,
    \beta
    \right)^*
    \lambda
    \; + \;
    \rho
    \,
    \|
    \alpha
    \, - \,
    \beta
    \|_2^2
    \right).
    \non
\eeq
Here, $\lambda \in \bbC^r$ is the vector of Lagrange multipliers,
$\rho$ is a positive parameter that introduces a quadratic penalty on
the deviation between $\alpha$ and $\beta$, and $\| \, \cdot \, \|_2$
is the Euclidean norm of a given vector. For $\rho = 0$, $\aL$
simplifies to the standard Lagrangian associated with the optimization
problem~(\ref{eq.Step1}).

\ei

ADMM is an iterative algorithm for minimization of the augmented
Lagrangian that consists of an $\alpha$-minimization step, a
$\beta$-minimization step, and a Lagrange multiplier update step
\begin{subequations}
  \label{eq.ADMM}
  \begin{align}
    \label{eq.F_update}
    \alpha^{k+1}
    \,& \DefinedAs \;
    \underset{\alpha}{\operatorname{arg \, min}}
    \;
    \aL
    \left( \alpha,\beta^k,\lambda^k \right),
    \\
    \label{eq.G_update}
    \beta^{k+1}
    \,& \DefinedAs \;
    \underset{\beta}{\operatorname{arg \, min}}
    \;
    \aL
    \left( \alpha^{k+1},\beta,\lambda^k \right),
    \\
    \label{eq.lambda_update}
    \lambda^{k+1}
    \,& \DefinedAs \;
    \lambda^{k}
    \,+\,
    \rho
    \left( \alpha^{k+1} \,-\, \beta^{k+1} \right).
  \end{align}
\end{subequations}
Starting with an initial point ($\beta^0,\lambda^0$), the iterations
are conducted until the desired feasibility tolerances,
$\epsilon_{\mathrm{prim}}$ and $\epsilon_{\mathrm{dual}}$, are met
\[
\| \alpha^{k+1} \, - \, \beta^{k+1} \|_2 \, \leq \,
\epsilon_{\mathrm{prim}} ~~~ \mbox{and} ~~~ \| \beta^{k+1} \, - \,
\beta^{k} \|_2 \, \leq \, \epsilon_{\mathrm{dual}}.
\]

In Appendix~\ref{sec.admm}, we utilize the respective structures of
the functions $J$ and $g$ in~\eqref{eq.Jell1-vec} and show that the
$\alpha$-minimization step amounts to solving an unconstrained
regularized quadratic program and that the solution to the
$\beta$-minimization step is obtained through the application of a 
soft-thresholding operator.

	\vspace*{-2ex}
\section{Examples}
	\vspace*{-1ex}
\label{sec.examples}

In this section, we apply sparsity-promoting DMD to three databases of
snapshots. The first database is obtained using a numerically
generated sequence of snapshots of the two-dimensional linearized
Navier--Stokes equations for plane Poiseuille flow with $Re = 10000$,
the second database results from an LES of a screeching supersonic
rectangular jet, and the third database contains the time-resolved
particle image velocimetry data of a flow through a cylinder
bundle.

	\vspace*{-2ex}
\subsection{Two-dimensional Poiseuille flow with $Re = 10000$}
	\vspace*{-1ex}
\label{sec.channel}

\begin{figure}
	\centering
	\includegraphics[width=0.75\columnwidth]{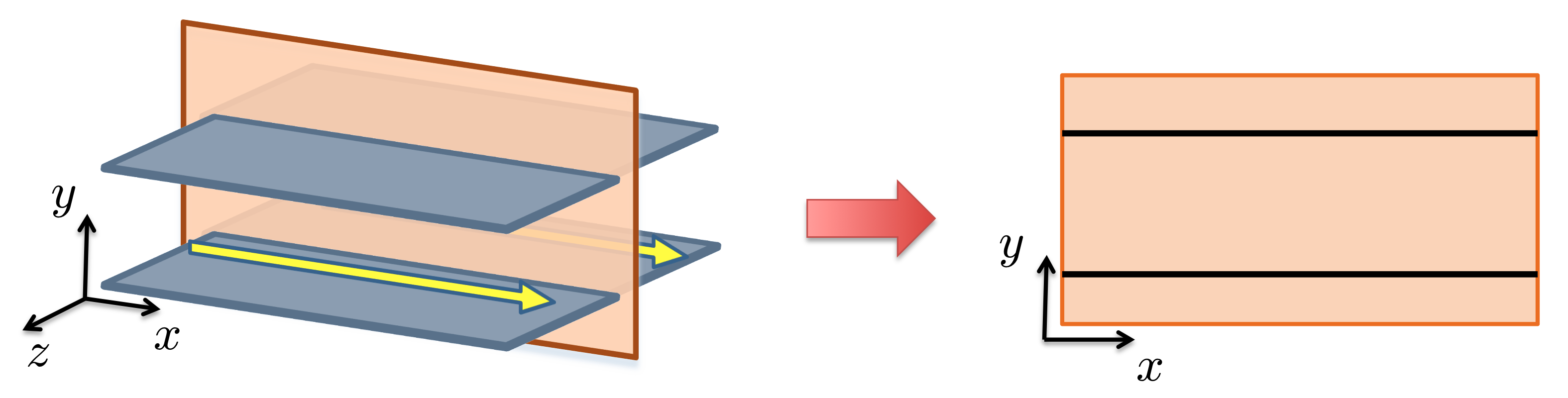}
	\caption{Geometry of a two-dimensional channel flow.}
	\label{fig.channel}
	\end{figure}

We begin illustration of the sparsity-promoting DMD algorithm by considering the two-dimensional linearized Navier--Stokes equations for plane Poiseuille flow with $Re = 10000$ (based on the centerline velocity and the channel half-width); see Figure~\ref{fig.channel} for geometry. The dynamics of the wall-normal velocity fluctuations with streamwise wavenumber $k_x$ are governed by the Orr-Sommerfeld equation~\cite{schhen01}. The discretized version of the linearized dynamics is obtained using a pseudo-spectral scheme with $M = 150$ collocation points in the wall-normal direction~\cite{weired00}, and the flow fluctuations with $k_x = 1$ are advanced in time using the matrix exponential with time step $\Delta t = 1$ and a randomly generated initial profile (that satisfies both homogeneous Dirichlet and Neumann boundary conditions). After a transient period of ten time-steps, $N = 100$ snapshots are taken to form the snapshot matrices, and we apply the standard DMD algorithm along with its sparsity-promoting variant.

\begin{figure}
  \centering \subfloat[] {\label{fig.channel-eigs}
    \begin{tikzpicture}
      \begin{axis}[ width=0.33\textwidth, height=0.275\textwidth,
        xlabel=$\operatorname{Re} \left( \log \, (\mu_i) \right)$,
        ylabel=$\operatorname{Im} \left( \log \, (\mu_i) \right)$,
        xmin = -0.77, xmax = 0.02, ymin = -1, ymax = -0.1 ]
        \addplot[only marks, mark size = 1.5pt, color=black, mark=o, line width = 0.65]
        table[x=real,y=imag] {data/channel/Eos.dat};
        \addplot[only marks, mark size = 1.5pt, color=red, mark=x, line width = 0.65]
        table[x=real,y=imag] {data/channel/Edmd.dat};
        \end{axis}
        \end{tikzpicture}
        }
        \hspace*{0.25cm}
         \subfloat[]
        {\label{fig.channel-amp-vs-fr}
        \begin{tikzpicture}
        \begin{axis}[
        width=0.33\textwidth, height=0.275\textwidth,
        	xlabel=$\operatorname{Im} \left( \log \, (\mu_i) \right)$,
	ylabel=$| \alpha_i |$,
	xmin = -1, xmax = -0.15,
	ymin = 0, ymax = 13
	]
        \addplot[only marks, mark size = 1.5pt, color=black, mark=o, line width = 0.65]
        	table[x=imag,y=amp] {data/channel/Amp_vs_re_fr.dat};
        \end{axis}
        \end{tikzpicture}
        }
        \hspace*{0.25cm}
         \subfloat[]
        {\label{fig.channel-amp-vs-real}
        \begin{tikzpicture}
        \begin{axis}[
        width=0.33\textwidth, height=0.275\textwidth,
        	xlabel=$\operatorname{Re} \left( \log \, (\mu_i) \right)$,
	ylabel=$| \alpha_i |$,
	xmin = -0.77, xmax = 0.05,
	ymin = 0, ymax = 13
	]
        \addplot[only marks, mark size = 1.5pt, color=black, mark=o, line width = 0.65]
        	table[x=real,y=amp] {data/channel/Amp_vs_re_fr.dat};
        \end{axis}
        \end{tikzpicture}
        }
        \caption{(a) Spectrum of the Orr-Sommerfeld operator (circles)
          along with the eigenvalues resulting from the standard DMD
          algorithm (crosses) for the two-dimensional Poiseuille flow
          with $Re = 10000$ and $k_x = 1$. Dependence of the absolute
          value of the DMD amplitudes $\alpha_i$ on (b) the frequency
          and (c) the real part of the corresponding DMD eigenvalues
          $\mu_i$.}
   \label{fig.channel-eigs-amp}
\end{figure}
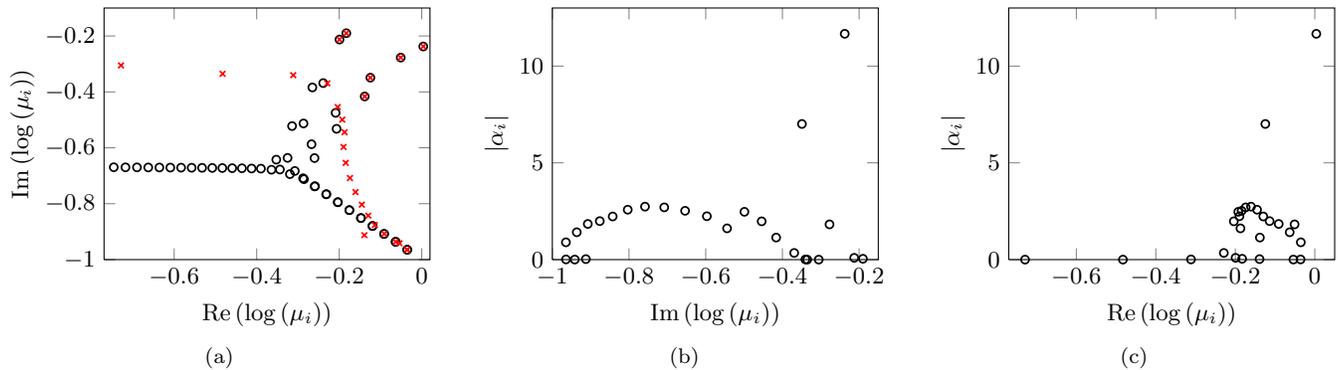

The spectrum of the Orr-Sommerfeld operator (circles) along with the
eigenvalues resulting from the standard DMD algorithm (crosses) are
shown in Figure~\ref{fig.channel-eigs}. The rank of the matrix of
snapshots $\Psi_0$ is $r = 26$, and the dependence of the absolute
value of the DMD amplitudes $\alpha_i$ on the frequency and the real
part of the corresponding DMD eigenvalues $\mu_i$ is displayed in
Figures~\ref{fig.channel-amp-vs-fr} and \ref{fig.channel-amp-vs-real},
respectively. As expected, the largest amplitude of DMD
modes corresponds to an unstable eigenvalue that generates
an exponentially growing Tollmien-Schlichting (TS) wave.

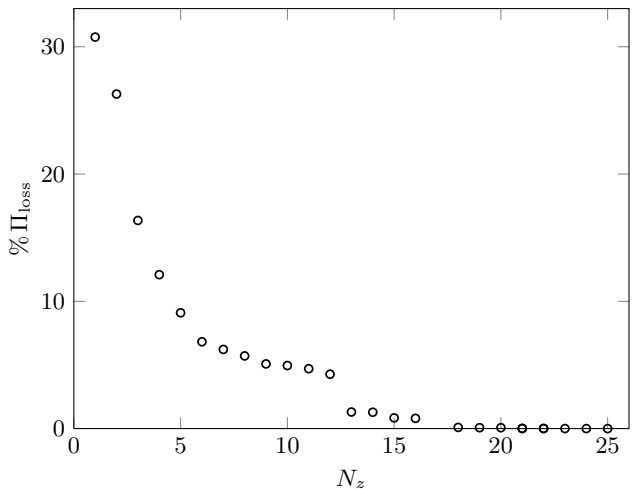
\begin{figure}
  \centering {
      \begin{tikzpicture}
        \begin{axis}[
        width=0.5\textwidth, height=0.4\textwidth,
        	xlabel=$N_z$,
	ylabel={$ \% \, \Ploss $},
	xmin = 0, xmax = 26,
	ymin = 0, ymax = 33
	]
        \addplot[only marks, mark size = 1.5pt, color=black, mark=o, line width = 0.65]
        	table[x=Nz,y=Ploss] {data/channel/posNzPloss.dat};
          \end{axis}
        \end{tikzpicture}
        }
        \caption{Performance loss
    	$ \% \, \Ploss \DefinedAs 100 \, \sqrt{J (\alpha) / J (0)} $
	of the optimal vector of amplitudes $\alpha$ resulting from the
    sparsity-promoting DMD algorithm (with $N_z$ DMD modes)
    for the Poiseuille flow example.}
     \label{fig.dmdsp-channel}
     \end{figure}

Figure~\ref{fig.dmdsp-channel} shows the residual $ \|
  \Psi_0 - \Phi \, \Dalpha \, \Vand \|_F $ of the optimal vector of
  amplitudes $\alpha$, resulting from the sparsity-promoting DMD
  algorithm, in fraction of $\| \Psi_0 \|_F$,
  \begin{equation}
    \% \, \Ploss
    \; \DefinedAs \;
    100 \, \sqrt{\dfrac{J (\alpha)}{J (0)}}
    \; = \;
    100 \, \dfrac{\| \Psi_0 - \Phi \, \Dalpha \, \Vand \|_F}{\| \Psi_0 \|_F},
    \non
\end{equation}
as a function of the number of the retained DMD modes, $N_z
\DefinedAs\hbox{{\bf{card}}} \left( \alpha \right)$. As our emphasis
on sparsity increases, a progressively smaller number of non-zero
elements in the vector $\alpha$ are obtained and the quality of the
least-squares approximation deteriorates. The
  sparsity-promoting DMD algorithm with $18$ modes provides nearly
  identical performance as the full-rank DMD algorithm. By reducing
  the number of modes to $13$, the least-squares residual gets
  compromised by only $1.3$ percent. While a reduction in the number
  of modes from $13$ to $12$ introduces performance degradation of
  approximately $3\%$, the reduction in the number of modes from $12$
  to $6$ degrades performance by only additional $2.5\%$. Further
  reduction in the number of DMD modes has a much more profound
  influence on the quality of approximation of numerically generated
  snapshots; for example, performance deterioration of almost $10\%$
  takes place when the number of DMD modes is reduced from $3$ to
  $2$. As we further discuss below, this sharp performance drop may be attributed to the fact that at least one fast, one slow, and one unstable mode has to be selected in order to capture the essential dynamical features of the original data sequence. The results displayed in Figure~\ref{fig.dmdsp-channel} suggest that a reasonable compromise between the quality of approximation (in the least-squares sense) and the number of modes in the Poiseuille flow example may be achieved with $N_z = 6$ DMD modes.
  
  	% Preamble: \pgfplotsset{compat=newest}

\begin{figure}
      \centering
      \subfloat[$N_z \, = \, 13$]
        {\label{fig.channel-spDMD13-eigs}
        \begin{tikzpicture}
        \begin{axis}[
        width=0.33\textwidth, height=0.275\textwidth,
	xlabel=$\operatorname{Re} \left( \log \, (\mu_i) \right)$,
	ylabel=$\operatorname{Im} \left( \log \, (\mu_i) \right)$,
	xmin = -0.77, xmax = 0.02,
	ymin = -1, ymax = -0.1
	]
        \addplot[only marks, mark size = 1.5pt, color=black, mark=o, line width = 0.65]
        table[x=real,y=imag] {data/channel/Edmd.dat};
        \addplot[only marks, mark size = 1.5pt, color=red, mark=x, line width = 0.65]
        table[x=real,y=imag] {data/channel/posEdmd13.dat};
        \end{axis}
        \end{tikzpicture}
        }
        \hspace*{0.25cm}
        \subfloat[$N_z \, = \, 3$]
        {\label{fig.channel-spDMD3-eigs}
        \begin{tikzpicture}
        \begin{axis}[
        width=0.33\textwidth, height=0.275\textwidth,
	xlabel=$\operatorname{Re} \left( \log \, (\mu_i) \right)$,
	%ylabel=$\operatorname{Im} \left( \log \, (\mu_i) \right)$,
	xmin = -0.77, xmax = 0.02,
	ymin = -1, ymax = -0.1
	]
        \addplot[only marks, mark size = 1.5pt, color=black, mark=o, line width = 0.65]
        table[x=real,y=imag] {data/channel/Edmd.dat};
        \addplot[only marks, mark size = 1.5pt, color=red, mark=x, line width = 0.65]
        table[x=real,y=imag] {data/channel/posEdmd3.dat};
        \end{axis}
        \end{tikzpicture}
        }
        \hspace*{0.25cm}    
        \subfloat[$N_z \, = \, 1$]
        {\label{fig.channel-spDMD1-eigs}
        \begin{tikzpicture}
        \begin{axis}[
        width=0.33\textwidth, height=0.275\textwidth,
	xlabel=$\operatorname{Re} \left( \log \, (\mu_i) \right)$,
	%ylabel=$\operatorname{Im} \left( \log \, (\mu_i) \right)$,
	xmin = -0.77, xmax = 0.02,
	ymin = -1, ymax = -0.1
	]
        \addplot[only marks, mark size = 1.5pt, color=black, mark=o, line width = 0.65]
        table[x=real,y=imag] {data/channel/Edmd.dat};
        \addplot[only marks, mark size = 1.5pt, color=red, mark=x, line width = 0.65]
        table[x=real,y=imag] {data/channel/posEdmd1.dat};
        \end{axis}
        \end{tikzpicture}
        }
        \\ 
         \subfloat[$N_z \, = \, 13$]
        {\label{fig.channel-spDMD13-amp}
        \begin{tikzpicture}
        \begin{axis}[
        width=0.345\textwidth, height=0.275\textwidth,
        	xlabel=$\operatorname{Im} \left( \log \, (\mu_i) \right)$,
	ylabel=$| \alpha_i |$,
	xmin = -1, xmax = -0.15,
	ymin = 0, ymax = 18
	]
        \addplot[only marks, mark size = 1.5pt, color=black, mark=o, line width = 0.65]
        	table[x=imag,y=amp] {data/channel/Amp_vs_re_fr.dat};
	\addplot[only marks, mark size = 1.5pt, color=red, mark=x, line width = 0.65]
        	table[x=imag,y=Amp] {data/channel/posEdmd13.dat};
        \end{axis}
        \end{tikzpicture}
        }
          \hspace*{0.25cm}
         \subfloat[$N_z \, = \, 3$]
        {\label{fig.channel-spDMD3-amp}
        \begin{tikzpicture}
        \begin{axis}[
        width=0.345\textwidth, height=0.275\textwidth,
        	xlabel=$\operatorname{Im} \left( \log \, (\mu_i) \right)$,
	%ylabel=$| \alpha_i |$,
	xmin = -1, xmax = -0.15,
	ymin = 0, ymax = 18
	]
        \addplot[only marks, mark size = 1.5pt, color=black, mark=o, line width = 0.65]
        	table[x=imag,y=amp] {data/channel/Amp_vs_re_fr.dat};
	\addplot[only marks, mark size = 1.5pt, color=red, mark=x, line width = 0.65]
        	table[x=imag,y=Amp] {data/channel/posEdmd3.dat};
        \end{axis}
        \end{tikzpicture}
        }
        \hspace*{0.25cm}
        \subfloat[$N_z \, = \, 1$]
        {\label{fig.channel-spDMD1-amp}
        \begin{tikzpicture}
        \begin{axis}[
        width=0.345\textwidth, height=0.275\textwidth,
        	xlabel=$\operatorname{Im} \left( \log \, (\mu_i) \right)$,
	%ylabel=$| \alpha_i |$,
	xmin = -1, xmax = -0.15,
	ymin = 0, ymax = 18
	]
        \addplot[only marks, mark size = 1.5pt, color=black, mark=o, line width = 0.65]
        	table[x=imag,y=amp] {data/channel/Amp_vs_re_fr.dat};
	\addplot[only marks, mark size = 1.5pt, color=red, mark=x, line width = 0.65]
        	table[x=imag,y=Amp] {data/channel/posEdmd1.dat};
        \end{axis}
        \end{tikzpicture}
        }
        \caption{Eigenvalues of $\Fdmd$ (first row) and the absolute values of
          the DMD amplitudes $\alpha_i$ (second row) for the
          Poiseuille flow example. The results are obtained using the
          standard DMD algorithm (circles) and the sparsity-promoting
          DMD algorithm (crosses) with $N_z$ DMD modes.}
  	 \label{fig.channel-spDMD}
	\end{figure}
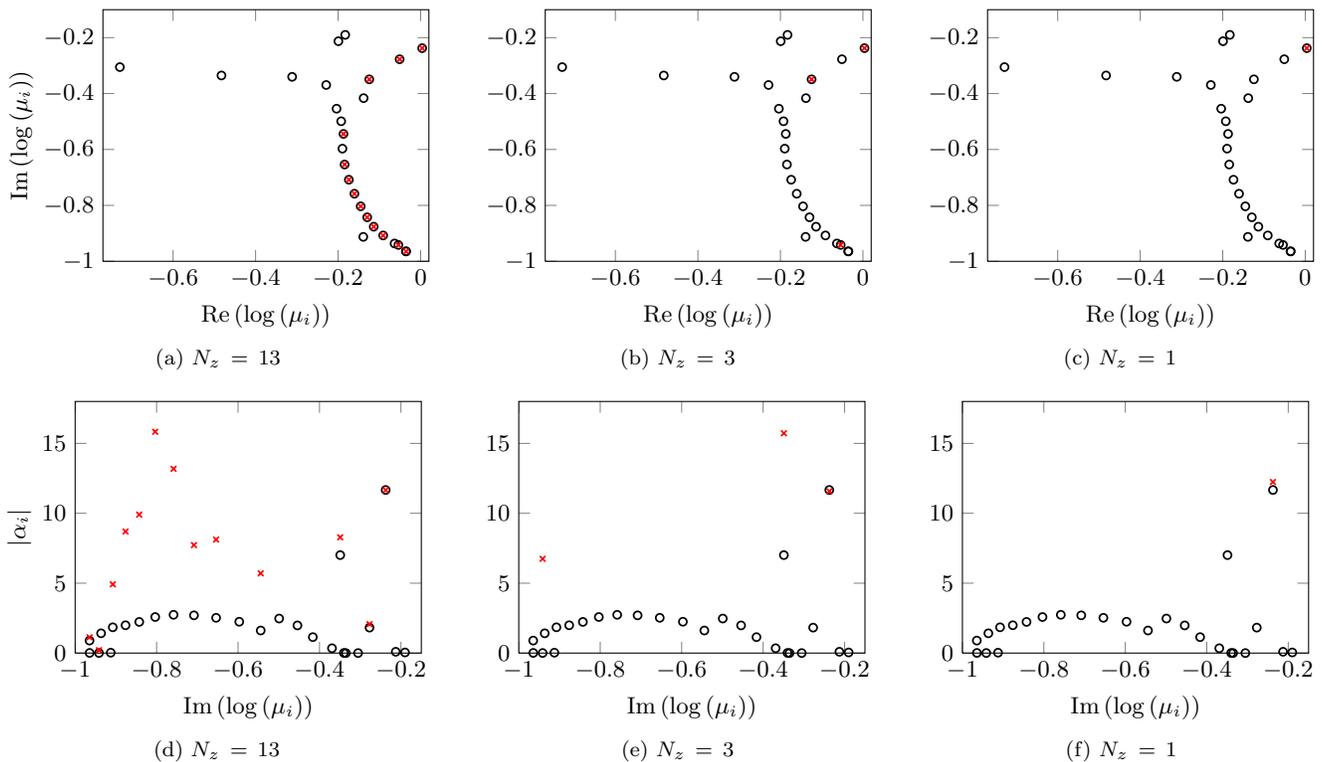

It is also instructive to assess how the sparsity-promoting procedure
selects the eigenvalues identified by the standard DMD algorithm. For
decreasing numbers of retained modes, Figure~\ref{fig.channel-spDMD}
illustrates this process both in the complex plane (first row) and
in the frequency-amplitude plane (second row). The DMD-spectrum for
plane Poiseuille flow consists of branches that describe distinct
features of the perturbation dynamics: the fast perturbation dynamics
in the center of the channel is captured by eigenvalues with phase
velocities $\operatorname{Re} \, (\log(\mu_i))/k_x$ that are larger
than the average base velocity (${\bar{U}}=2/3$), and the slower
dynamics which takes place near the channel walls is described by
eigenvalues with smaller phase velocities. As our emphasis on sparsity
increases, we observe a selection of modes from both the slow and fast
branches. While for $N_z=13$ modes nearly the entire slow and fast
branches (equivalent to the A- and P-branches of plane Poiseuille
flow~\cite{schhen01}) are selected, a progressive coarsening on each branch is seen
for a smaller number of modes. This process continues until, for
$N_z=3$, only one fast, one slow, and one unstable mode is chosen to
represent the dynamics contained in the original data
sequence. Finally, in the limit of only a single mode, for $N_z=1$,
the unstable TS wave is selected by the sparsity-promoting DMD algorithm.

% Preamble: \pgfplotsset{compat=newest}
     
     % Preamble: \pgfplotsset{compat=newest}

	\vspace*{-2ex}
\subsection{A screeching supersonic jet}
	\vspace*{-1ex}
\label{sec.screech}

Screech is a component of supersonic jet noise that is connected to
the presence of a train of shock cells within the jet
column~\cite{tam95}.  For a turbulent jet, the unsteady shear layers
interact with the shocks to create sound.  While this process is
generally broadbanded, screech is a special case of shock-noise that
arises from the creation of a feedback loop between the
upstream-propagating part of the acoustic field and the generation of
new disturbances at the nozzle lip. This self-sustaining feedback loop
leads to an extremely loud (narrow-banded) screech tone at a specific
fundamental frequency. The presence of a tonal process embedded in an
otherwise broadbanded turbulent flow makes the screeching jet an
excellent test case for the sparsity-promoting DMD method. The
objective of the developed method is to extract the entire coherent
screech feedback loop from the turbulent data and to describe the
screech mechanism with as few modes as possible.

\begin{figure}
  \centering
   \subfloat[] {
    \label{fig.screech_geometry}
    \includegraphics[width=0.33\textwidth, height=0.275\textwidth]
    {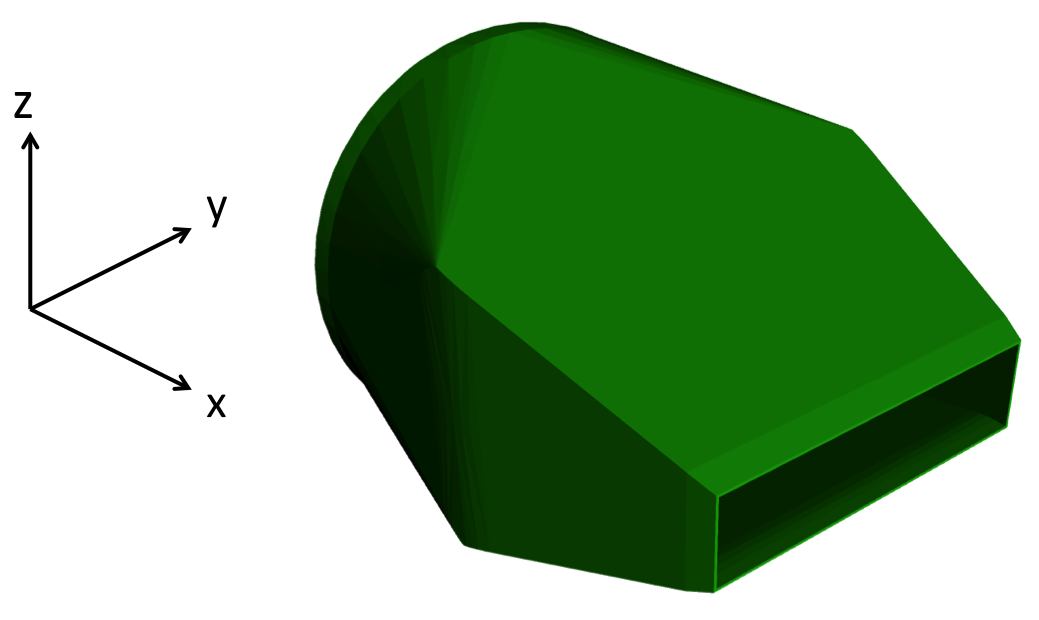} } 
   \hspace*{0.5cm}
    \subfloat[] {
    \label{fig.screech_T}
    \includegraphics[width=0.5\textwidth, height=0.275\textwidth]{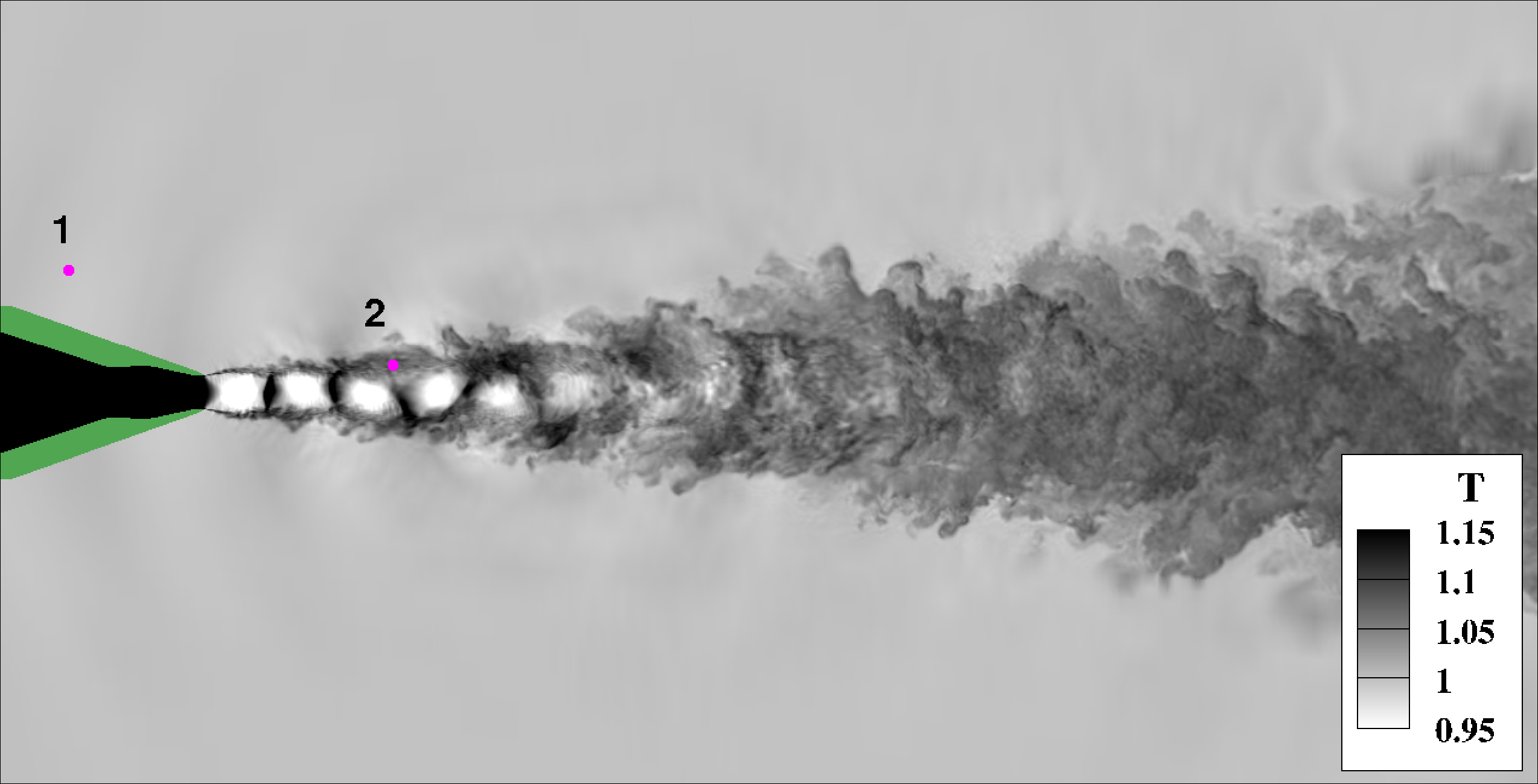}
    }
  \caption{(a) Rectangular nozzle geometry of aspect ratio 4. (b) Temperature contours in the ($x,z$)-centerplane taken from a snapshot of the screeching jet.  Time
    histories of pressure were recorded at probe locations indicated
    by circles $1$ and $2$.}
\end{figure}

The supersonic jet used in this example was produced by a convergent
rectangular nozzle of aspect ratio 4, precisely matching the geometry
of an experimental nozzle~\cite{frate11}; see Figure~\ref{fig.screech_geometry} for geometry. The entire flow inside,
outside, and downstream of the nozzle was simulated using the
low-dissipation low-dispersion LES solver \texttt{charles} on an
unstructured mesh with about 45 million control volumes. This
simulation was a part of a sequence involving different mesh
resolutions, and it was validated against the experimental
measurements~\cite{nichols11}.

% Preamble: \pgfplotsset{compat=newest}
\begin{figure}
  \centering {
      \begin{tikzpicture}
        \begin{loglogaxis}[
        width=0.5\textwidth, height=0.4\textwidth,
        	xlabel=$St$,
	xmin = 0.0736, xmax = 9.3592,
	ymin = 2e-6, ymax = 5.e2,
	ytick={1.e-4,1.e-2,1.e0,1.e2}
	]
        \addplot[mark size = 1.6pt, color=black, mark=x, line width = 0.65]
        	table[x=St,y=p1] {data/jet/p1psd.dat};
         \addplot[mark size = 1.6pt, color=black, mark=o, line width = 0.65]
        	table[x=St,y=p2] {data/jet/p2psd.dat};
        \end{loglogaxis}
        \end{tikzpicture}
        }
        \caption{Power spectra of pressure corresponding to the
          measurement locations indicated in
          Figure~\ref{fig.screech_T}.  While both spectra peak at $St
          \approx 0.3$, the spectrum at location 2 (circles) is more
          broadbanded than at location 1 (crosses).}
        \label{fig.probes}
\end{figure}
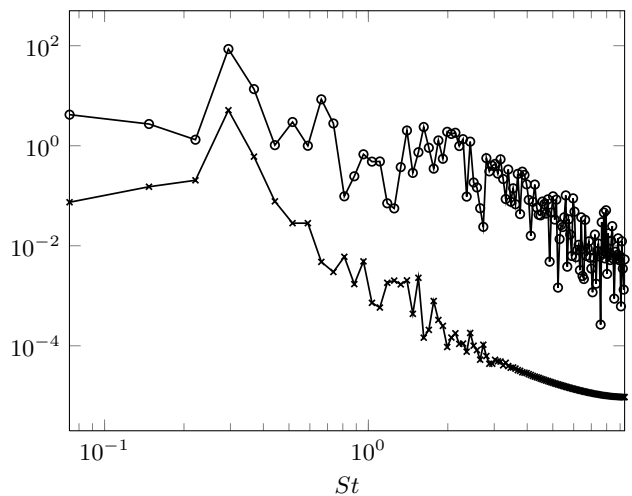

The stagnation pressure and temperature inside the nozzle were set so
that the jet Mach number $M_j = 1.4$ and the fully expanded jet
temperature matched the ambient temperature. A rectangular nozzle
whose interior cross section area was decreasing monotonically from
inlet to exit was used. Since the nozzle did not have a diverging
section before its exit, the flow left the nozzle in an under-expanded
(sonic) state and continued to expand downstream to reach the
supersonic fully-expanded condition. This induced a train of
diamond-shaped shock cells as shown in Figure~\ref{fig.screech_T}. The
figure shows contours of temperature on a centerplane cross section
taken through the narrow dimension of the nozzle.  From an animation
of the jet, which is available at \texttt{www.umn.edu/$\sim$mihailo/software/dmdsp/screech/}, it can be observed that the first two shock cells are almost stationary, but the third
and fourth shock cells undergo transverse oscillations along the
narrow (vertical) dimension of the jet.  The transverse oscillation of
the shock cells occurs precisely at the screech frequency of the jet,
and is connected to a strong upstream-oriented acoustic tone.

Figure~\ref{fig.probes} shows the spectra of the pressure recorded at
the locations indicated by the circles in
Figure~\ref{fig.screech_T}. At location~1, which coincides with the
center of the upstream-directed acoustic beam associated with the
screech tone, the spectrum is relatively narrow-banded. Since the
sample window contains approximately four oscillations of the screech
tone, the fourth Fourier coefficient has the largest amplitude. In
contrast to location~1, location~2 was taken in the center of one of
the turbulent shear layers. The turbulent fluctuations induce a broad
range of frequencies in the corresponding spectrum (circles in
Figure~\ref{fig.probes}). At both locations, spectra show a strong
peak at Strouhal number $St \approx 0.3$ (based on the fully expanded
jet velocity and the nozzle equivalent diameter). This peak
corresponds to the screech tone frequency predicted for a Mach $1.4$
jet from a 4:1 aspect ratio rectangular nozzle~\cite{tam95,nichols11}.

The database used for the DMD analysis consisted of 257 snapshots (so
that $N = 256$) of the full three-dimensional pressure and velocity
fields.  The snapshots were equispaced in time with an interval of
$\Delta t = 0.0528 D_e / u_j$, where $D_e$ is the nozzle equivalent
diameter (the diameter of the circle of same area as the nozzle exit)
and $u_j$ is the fully expanded jet velocity. Although the
computational domain for the LES extended approximately $32 D_e$
downstream of the nozzle exit, DMD was applied to a subdomain focusing
on the shock cells within the jet's potential core and the surrounding
acoustic field (this domain extends to $10 D_e$).  This restriction
reduced the number of cells from $45$ million to $8$ million. In spite
of this reduction, each snapshot required 256Mb of storage in double
precision format. To handle such large matrices, DMD was implemented
using a MapReduce framework so that the matrix could be stored and
processed across several storage discs. In particular, the algorithm
relied upon a MapReduce QR-factorization of tall-and-skinny matrices
developed in~\cite{constantine11}.

\begin{figure}
  \centering 
  \subfloat[] {\label{fig.jetA_vs_w}
    \begin{tikzpicture}
      \begin{axis}[ width=0.33\textwidth, height=0.275\textwidth,
        xlabel=$\operatorname{Im} \left( \log \, (\mu_i) \right)$,
        ylabel=$| \alpha_i |$, xmin = -3.2, xmax = 3.2, ymin = 0, ymax
        = 36000 ]
        \addplot[only marks, mark size = 1.5pt, color=black, mark=o, line width = 0.65]
        	table[x=w,y=Amp] {data/jet/jetA_vs_w.dat};
        \end{axis}
        \end{tikzpicture}
        }
        \hspace*{0.25cm}
         \subfloat[]
        {\label{fig.jetA_vs_Re}
        \begin{tikzpicture}
        \begin{axis}[
        width=0.33\textwidth, height=0.275\textwidth,
        	xlabel=$\operatorname{Re} \left( \log \, (\mu_i) \right)$,
	%ylabel=$| \alpha_i |$,
	xmin = -0.8, xmax = 0,
	ymin = 0, ymax = 36000
	]
        \addplot[only marks, mark size = 1.5pt, color=black, mark=o, line width = 0.65]
        	table[x=real,y=Amp] {data/jet/jetA_vs_Re.dat};
        \end{axis}
        \end{tikzpicture}
        }
        \hspace*{0.25cm}
        \subfloat[]
        {\label{fig.jetA_vs_Re_zoom}
        \begin{tikzpicture}
        \begin{axis}[
        width=0.33\textwidth, height=0.275\textwidth,
        	xlabel=$\operatorname{Re} \left( \log \, (\mu_i) \right)$,
	%ylabel=$| \alpha_i |$,
	xmin = -0.028, xmax = 0,
	ymin = 0, ymax = 2600
	]
        \addplot[only marks, mark size = 1.5pt, color=black, mark=o, line width = 0.65]
        	table[x=real,y=Amp] {data/jet/jetA_vs_Re.dat};
        \end{axis}
        \end{tikzpicture}
        }
        \caption{Dependence of the absolute value of the DMD
          amplitudes $\alpha_i$ on (a) the frequency and (b,c) the
          real part of the corresponding DMD eigenvalues $\mu_i$ for
          the screeching jet example. Subfigure~(c) represents a
          zoomed version of subfigure~(b) and it focuses on the
          amplitudes that correspond to lightly damped eigenvalues.}
    \label{fig.dmd-screech-amplitude-vs-fr-Re}
\end{figure}
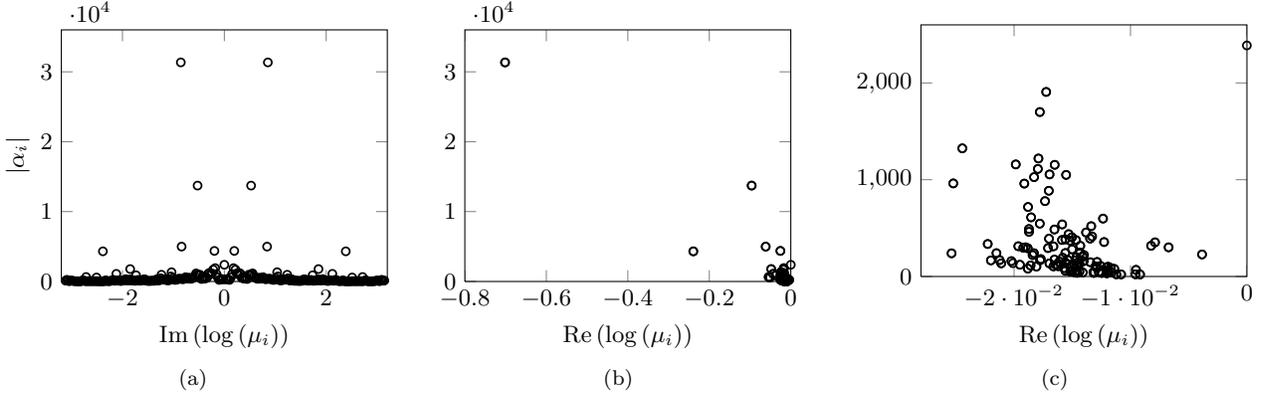

Figure~\ref{fig.jetA_vs_w} illustrates the frequency dependence of the
absolute value of the amplitudes of the DMD modes obtained by solving
the optimization problem~(\ref{eq.alpha_opt}). We note that it is not trivial to
identify, by mere inspection, a subset of DMD modes that has the
strongest influence on the quality of the least-squares
approximation. As shown in Figure~\ref{fig.jetA_vs_Re}, the largest
amplitudes originate from eigenvalues that are strongly damped. In
what follows, we demonstrate that keeping only a subset of modes with
largest amplitudes can lead to poor quality of approximation of
numerically generated snapshots.

% Preamble: \pgfplotsset{compat=newest}

\begin{figure}
  \centering \subfloat[]
  {\label{fig.dmdsp-screech-Nz-vs-gamma}
        \begin{tikzpicture}
        \begin{semilogxaxis}[
        width=0.45\textwidth, height=0.33\textwidth,
        	xlabel=$\gamma$,
	ylabel=$\hbox{{\bf{card}}} \, (\alpha)$,
	xmin = 110, xmax = 33000,
	ymin = 0, ymax = 280
	]
        \addplot[only marks, mark size = 1pt, color=black, mark=o]
        	table[x=gamma,y=Nz] {data/jet/jetJnz.dat};
        \end{semilogxaxis}
        \end{tikzpicture}
        }
      \hspace*{0.5cm}
      \subfloat[]
      {\label{fig.dmdsp-screech-sqrtFrob-vs-gamma}
        \begin{tikzpicture}
        \begin{semilogxaxis}[
        width=0.45\textwidth, height=0.33\textwidth,
        	xlabel=$\gamma$,
	ylabel=$ \% \, \Ploss$,
	xmin = 110, xmax = 33000,
	ymin = 0, ymax = 17.5
	]
        \addplot[only marks, mark size = 1pt, color=black, mark=o]
        	table[x=gamma,y=Ploss] {data/jet/jetPloss.dat};
        \end{semilogxaxis}
        \end{tikzpicture}
      }\caption{(a) The sparsity level $\hbox{{\bf{card}}} \, (\alpha)$ and 
      (b) the performance loss $ \% \, \Ploss \DefinedAs 100 \, \sqrt{J (\alpha) / J (0)} $ 
      of the optimal vector of amplitudes $\alpha$ resulting from the sparsity-promoting DMD algorithm for the screeching jet example.}
     \label{fig.dmdsp-screech-Nz_sqrtFrob_polished-vs-gamma}
\end{figure}
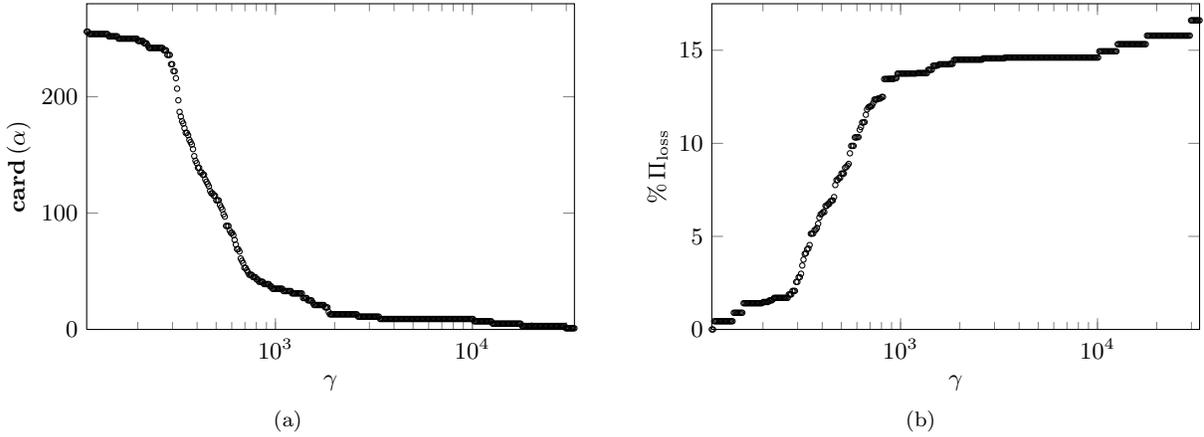

The sparsity level $\hbox{{\bf{card}}} \, (\alpha)$ and the
performance loss $ \% \, \Ploss \DefinedAs 100 \, \sqrt{J (\alpha) / J (0)} $ for the optimal vector of amplitudes $\alpha$ resulting from the sparsity-promoting DMD algorithm are shown
in Figure~\ref{fig.dmdsp-screech-Nz_sqrtFrob_polished-vs-gamma} as a
function of the user-specified parameter $\gamma$ (a measure of
preference between approximation quality and solution sparsity). As
expected, larger values of $\gamma$ encourage sparser solutions, at
the expense of compromising quality of the least-squares
approximation. The values of $\gamma$ in
Figure~\ref{fig.dmdsp-screech-Nz_sqrtFrob_polished-vs-gamma} are
selected in such a way that $\gamma_{\min}$ induces a dense vector
$\alpha$ (with $256$ non-zero elements), and $\gamma_{\max}$ induces
$\alpha$ with a single non-zero element.

% Preamble: \pgfplotsset{compat=newest}

\begin{figure}
  \centering \subfloat[$N_z \, = \, 47$] {\label{fig.jetAsp_vs_w47}
    \begin{tikzpicture}
      \begin{semilogyaxis}[ width=0.33\textwidth,
        height=0.275\textwidth, xlabel=$\operatorname{Im} \left( \log
          \, (\mu_i) \right)$, ylabel=$| \alpha_i |$, xmin = -3.5,
        xmax = 3.5, ymin = 10, ymax = 5.e4 ]
        \addplot[only marks, mark size = 1.5pt, color=black, mark=o, line width = 0.65]
        	table[x=w,y=Amp] {data/jet/jetA_vs_w.dat};
	\addplot[only marks, mark size = 1.5pt, color=red, mark=x, line width = 0.65]
        	table[x=w,y=Amp] {data/jet/jetAsp_vs_w47.dat};
        \end{semilogyaxis}
        \end{tikzpicture}
        }
        \hspace*{0.25cm}
         \subfloat[$N_z \, = \, 5$]
        {\label{fig.jetAsp_vs_w5}
        \begin{tikzpicture}
        \begin{axis}[
        width=0.33\textwidth, height=0.275\textwidth,
        	xlabel=$\operatorname{Im} \left( \log \, (\mu_i) \right)$,
	% ylabel=$| \alpha_i |$,
	xmin = -0.26, xmax = 0.26,
	ymin = 0, ymax = 5000
	]
        \addplot[only marks, mark size = 1.5pt, color=black, mark=o, line width = 0.65]
        	table[x=w,y=Amp] {data/jet/jetA_vs_w.dat};
	\addplot[only marks, mark size = 1.5pt, color=red, mark=x, line width = 0.65]
        	table[x=w,y=Amp] {data/jet/jetAsp_vs_w5.dat};
        \end{axis}
        \end{tikzpicture}
        }
       \hspace*{0.25cm}
	 \subfloat[$N_z \, = \, 3$]
        {\label{fig.jetAsp_vs_w3}
        \begin{tikzpicture}
        \begin{axis}[
        width=0.33\textwidth, height=0.275\textwidth,
        xlabel=$\operatorname{Im} \left( \log \, (\mu_i) \right)$,
	%ylabel=$| \alpha_i |$,
	xmin = -0.26, xmax = 0.26,
	ymin = 0, ymax = 5000
	]
        \addplot[only marks, mark size = 1.5pt, color=black, mark=o, line width = 0.65]
        	table[x=w,y=Amp] {data/jet/jetA_vs_w.dat};
	\addplot[only marks, mark size = 1.5pt, color=red, mark=x, line width = 0.65]
        	table[x=w,y=Amp] {data/jet/jetAsp_vs_w3.dat};
        \end{axis}
        \end{tikzpicture}
        }
        \caption{Dependence of the absolute value of the amplitudes
          $\alpha_i$ on the frequency (imaginary part) of the
          corresponding eigenvalues $\mu_i$ for the screeching jet
          example. The results are obtained using the standard DMD
          algorithm (circles) and the sparsity-promoting DMD algorithm
          (crosses) with $N_z$ DMD modes.}
        \label{dmd-vs-dmdsp-screech-amplitude-vs-fr-Re}
\end{figure}
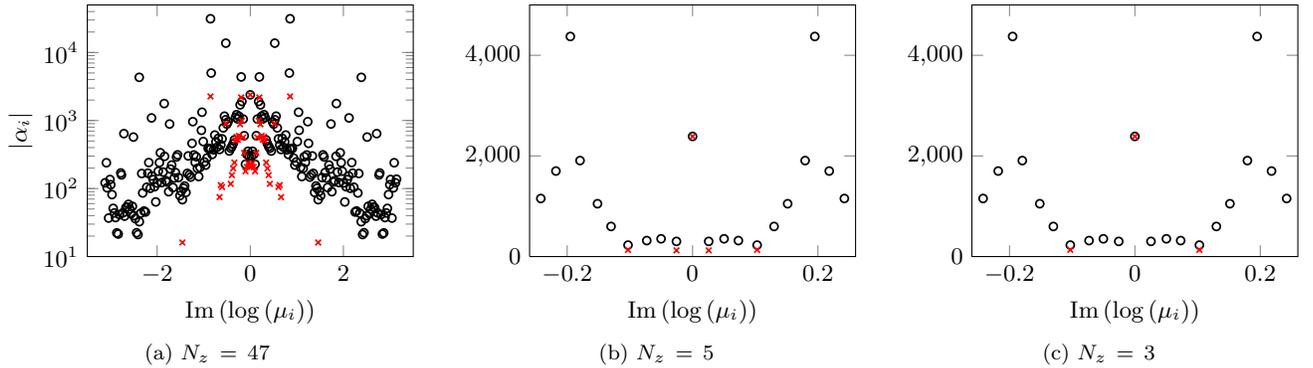

Eigenvalues resulting from the standard DMD algorithm (circles) along
with the subset of $N_z$ eigenvalues selected by the
sparsity-promoting DMD algorithm (crosses) are shown in
Figure~\ref{fig.Edmd-Edmdsp-screech}. Eigenvalues in the interior of
the unit circle are strongly damped. Since a strongly damped mode
influences only early stages of time evolution, the associated
amplitude $|\alpha_i|$ can be large; see Figure~\ref{fig.jetA_vs_Re}
for an illustration. Rather than focusing only on the modes with
largest amplitudes, the sparsity-promoting DMD identifies modes that
have the strongest influence on the entire time history of available
snapshots. While the selection of the retained eigenvalues is
non-trivial, it increasingly concentrates on the low-frequency modes
as $N_z$ decreases. For $N_z = 3,$ only the mean flow and one dominant
frequency pair (that corresponds to the fundamental frequency of the
screech tone) remain, while for $N_z = 5$ a second, lower frequency is
identified. In this low-$N_z$ limit we do see that the
sparsity-promoting DMD approximates the data sequence using the most
prevalent structures. The corresponding amplitudes for the various
truncations $N_z$ are displayed in
Figure~\ref{dmd-vs-dmdsp-screech-amplitude-vs-fr-Re}. Again, as $N_z$
decreases a concentration on low frequencies is observed. We also note
that the amplitudes of the original DMD modes do not provide
sufficient guidance in reducing the full set of DMD modes to a few
relevant ones.

% Preamble: \pgfplotsset{compat=newest}
\begin{figure}
  \centering \subfloat[$N_z \, = \, 47$] {\label{fig.jetEdmd47}
    \begin{tikzpicture}
        \begin{axis}[
        width=0.33\textwidth, height=0.33\textwidth,
        	xlabel=$\operatorname{Re} \left( \mu_i \right)$,
         ylabel=$\operatorname{Im} \left( \mu_i \right)$,
	xmin = -1, xmax = 1,
	ymin = -1, ymax = 1
	]
        \addplot[only marks, mark size = 1.5pt, color=black, mark=o]
        	table[x=real,y=imag]{data/jet/jetEdmd.dat};
        \addplot[only marks, mark size = 1.5pt, color=red, mark=x]
        	table[x=real,y=imag]{data/jet/jetEdmd47.dat};
        \end{axis}
        \end{tikzpicture}
        }
         \hspace*{0.05cm}
        \subfloat[$N_z \, = \, 5$]
        {\label{fig.jetEdmd5}
        \begin{tikzpicture}
        \begin{axis}[
        width=0.33\textwidth, height=0.33\textwidth,
        	xlabel=$\operatorname{Re} \left( \mu_i \right)$,
	%ylabel=$\operatorname{Im} \left( \mu_i \right)$,
	xmin = 0.92 , xmax = 1,
	ymin = -0.2, ymax = 0.2
	]
        \addplot[only marks, mark size = 1.5pt, color=black, mark=o]
        	table[x=real,y=imag]{data/jet/jetEdmd.dat};
        \addplot[only marks, mark size = 1.5pt, color=red, mark=x]
        	table[x=real,y=imag]{data/jet/jetEdmd5.dat};
	\addplot[no markers, line width = 1.25, color=blue, dashed]
        	table[x=real,y=imag]{data/jet/circle.dat};
        \end{axis}
        \end{tikzpicture}
        }
         \hspace*{0.05cm}
        \subfloat[$N_z \, = \, 3$]
        {\label{fig.jetEdmd3}
        \begin{tikzpicture}
        \begin{axis}[
        width=0.33\textwidth, height=0.33\textwidth,
        	xlabel=$\operatorname{Re} \left( \mu_i \right)$,
	%ylabel=$\operatorname{Im} \left( \mu_i \right)$,
	xmin = 0.92 , xmax = 1,
	ymin = -0.2, ymax = 0.2
	]
        \addplot[only marks, mark size = 1.5pt, color=black, mark=o]
        	table[x=real,y=imag]{data/jet/jetEdmd.dat};
        \addplot[only marks, mark size = 1.5pt, color=red, mark=x]
        	table[x=real,y=imag]{data/jet/jetEdmd3.dat};
	\addplot[no markers, line width = 1.25, color=blue, dashed]
        	table[x=real,y=imag]{data/jet/circle.dat};
        \end{axis}
        \end{tikzpicture}
        }
        \caption{Eigenvalues resulting from the standard DMD algorithm
          (circles) along with the subset of $N_z$ eigenvalues
          selected by the sparsity-promoting DMD algorithm (crosses)
          for the screeching jet example. In subfigures (b) and (c),
          the dashed curves identify the unit circle.}
    \label{fig.Edmd-Edmdsp-screech}
\end{figure}
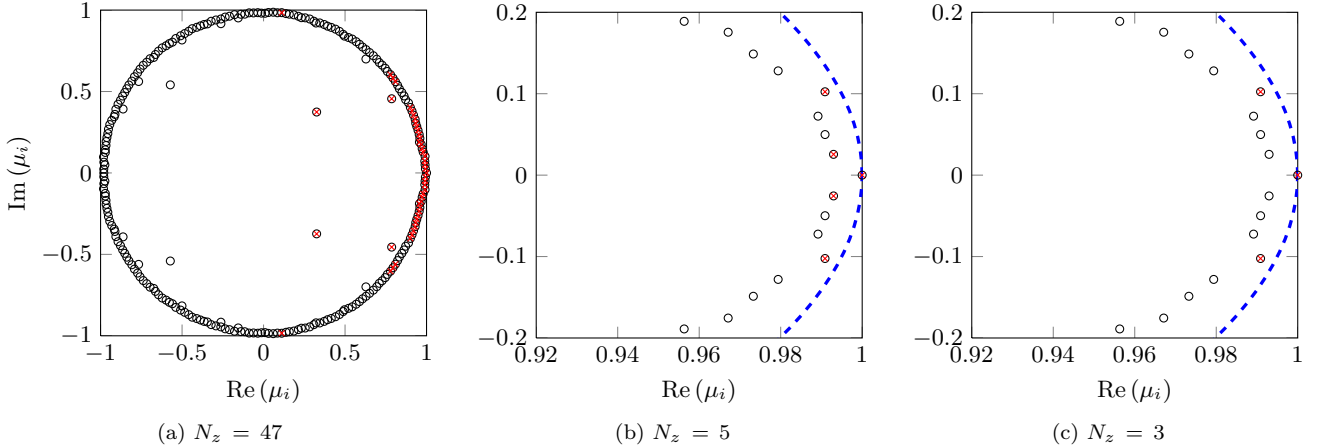

% Preamble: \pgfplotsset{compat=newest}

\begin{figure}
  \centering {
    \begin{tikzpicture}
        \begin{axis}[
        width=0.5\textwidth, height=0.4\textwidth,
        	xlabel=$N_z$,
	ylabel={$ \% \, \Ploss$},
                xmin = 0, xmax = 260, ymin = 0, ymax = 17.5 ]
        \addplot[only marks, mark size = 1.5pt, color=black, mark=o, line width = 0.65]
        	table[x=Nz,y=Ploss] {data/jet/jetspPloss_vs_Nz.dat};
%%         \addplot[no markers, line width = 1, color=black]
%%        	table[x=Nz,y=Ploss] {data/jet/jetlowPloss_vs_Nz.dat};
        \end{axis}
        \end{tikzpicture}
        }
        \caption{The optimal performance loss, $ \% \, \Ploss \DefinedAs 100 \, \sqrt{J (\alpha) / J (0)} $, of the sparsity-promoting DMD algorithm for the screeching jet example as a function of the number of retained modes.}
     \label{fig.dmdsp-vs-dmdlr}
\end{figure}
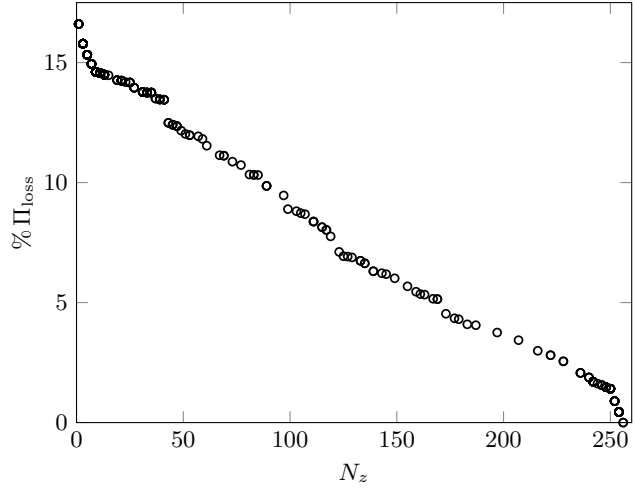

Figure~\ref{fig.dmdsp-vs-dmdlr} illustrates performance of the
sparsity-promoting DMD algorithm developed in this paper in terms of
the number of identified structures. The performance is quantified by
the Frobenius norm of the approximation error between a
low-dimensional representation and the full data sequence in fraction
of the Frobenius norm of the full data sequence. As expected, the
performance of our algorithm improves as the number of modes is
increased. For the screeching jet example, the sparsity-promoting DMD
with $N_z=3$ gives $St = 0.3104$ which agrees well with the frequency
$St=0.3$ measured from a sequence containing $10$ times the number of
snapshots.

% Preamble: \pgfplotsset{compat=newest}

\begin{figure}
  \begin{center}
    \includegraphics[width=0.75\textwidth]{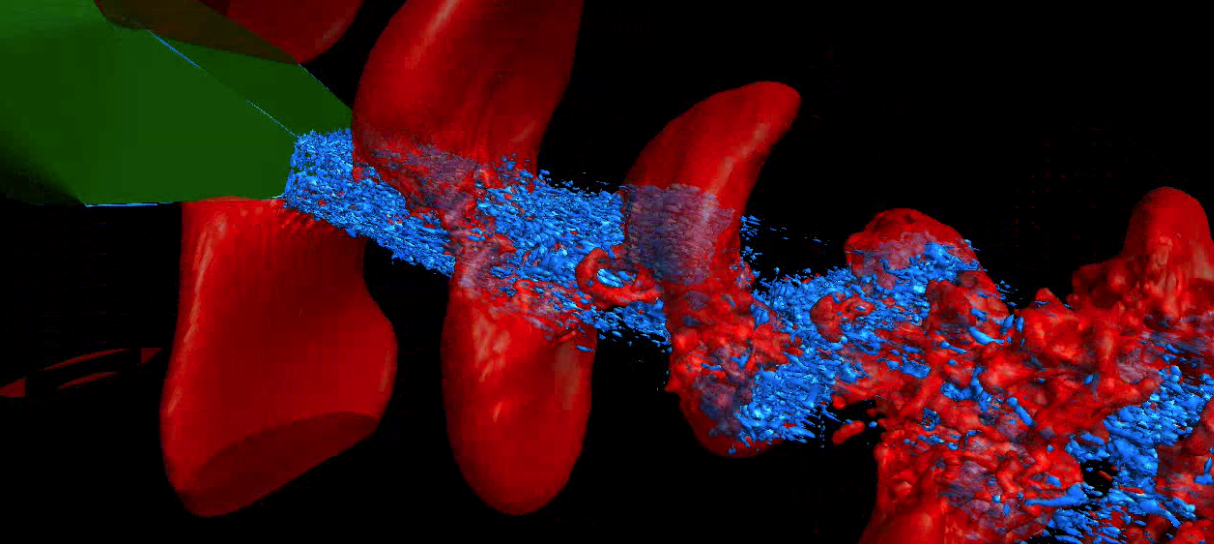}
  \end{center}
  \caption{\label{fig.screech-mode} The three-dimensional DMD mode
    associated with the dominant frequency pair (that corresponds to
    the fundamental frequency of the screech tone, $St \approx 0.3$).  A red isosurface
    of perturbation pressure is shown together with a blue isosurface
    of the perturbation dilatation. An animation of this flow
    field is available at \texttt{www.umn.edu/$\sim$mihailo/software/dmdsp/screech/}.}
\end{figure}

Finally, Figure~\ref{fig.screech-mode} visualizes the
three-dimensional mode associated with the dominant frequency pair
resulting from the sparsity-promoting DMD algorithm. Isosurfaces of
pressure and dilatation fluctuations are identified by red and blue
colors, respectively. An animation of this mode reveals that
the dilatation corresponds to a flapping mode along the narrow
dimension of the jet; please see \texttt{www.umn.edu/$\sim$mihailo/software/dmdsp/screech/}. As the shock cells oscillate transversely, they encroach into the jet shear layers, and compress the oncoming jet fluid in a region where previously there was no compression. This
time-periodic compression is connected to the upstream propagating
acoustic wave at the exterior of the jet. Furthermore, the animation
of the DMD mode reveals that -- close to the jet -- the acoustic wave
does not propagate smoothly in the upstream direction. Instead, the
acoustic wave hops from one spot of negative dilatation to the next as
it proceeds upstream.  As the oscillating shock cells penetrate into
the jet shear layers, they are also rotated by the shear. When this
happens, the outer tip of the shock cell sweeps from downstream to
upstream, providing an impulse to the exterior acoustic perturbation
which ``kicks'' it upstream to the next shock cell.

	\vspace*{-2ex}
\subsection{Flow through a cylinder bundle}
	\vspace*{-1ex}
\label{sec.cylinder}

Flow through cylinders in a bundle configuration is often encountered in the utility and energy conversion industry. Owing to their versatility and efficiency~\cite{kakac:97}, cross-flow heat exchangers account for the vast majority of heat exchangers in oil refining, process engineering, petroleum extraction, and power generation sectors. Consequently, even a modest improvement in their effectiveness and operational margins would have a significant impact on production efficiency. In spite of their widespread use, the details of the flow (and heat transfer) through cylinder bundles are far from fully understood, thereby making this type of flow, along with its simplified variants, the subject of active research~\cite{rollet-miet:1999,sumner:2000,benhamadouche:2003,moulinec:2004}.

Vortex shedding and complex wake interactions can induce vibrations
and structural resonances which, in turn, may result in fretting wear,
collisional damage, material fatigue, creep, and ultimately in
cracking. Even though the flow through a cylinder bundle is very complex, it is characterized by well-defined shedding frequencies~\cite{liang:2007}. For single-row cylinders, only a few distinct frequencies are detected, while for more complex array configurations, a multitude of precise shedding frequencies are observed. The presence of distinct shedding frequencies makes this type of flow well-suited for a decomposition of the flow fields into single-frequency modes.

In addition to the multiple-frequency behavior, a characteristic oscillatory pattern -- labelled as a ``flip-flop phenomenon'' -- is typically observed in the exit stream of the flow through cylinder bundles~\cite{simonin,hassan:1999}. The observed oscillatory pattern consists of a meta-stable deflection of the exiting jet off the centerline location. A statistical analysis of the velocity field in a cross-stream section shows a strongly bimodal distribution. Over a wide range of Reynolds numbers, the Strouhal number associated with the oscillation between these two states is approximately Reynolds-number-independent. This flow feature can also be represented by a dynamic mode decomposition of the flow fields.

\begin{figure}
  \centering 
  \subfloat[] {
    \label{fig.cyl_geometry}
    \includegraphics[width=0.275\textwidth, height=0.275\textwidth]
    {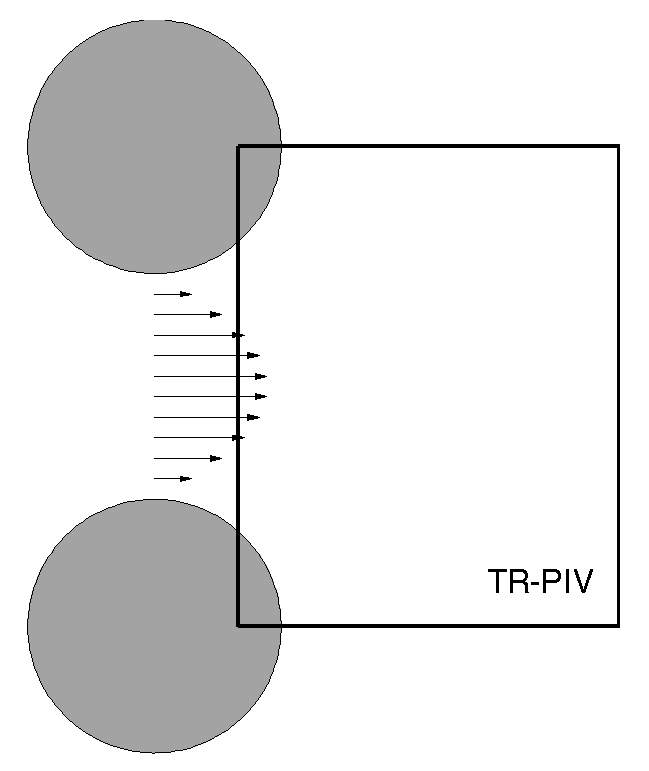} } 
    \subfloat[] {
    \label{fig.cyl_Snapshots_a}
    \includegraphics[width=0.33\textwidth, height=0.275\textwidth]
    {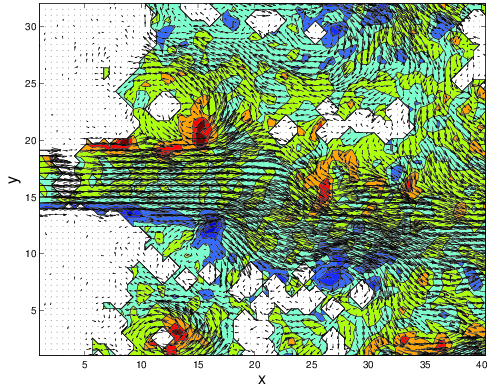} }
  % \hspace*{0.25cm}
  \subfloat[] {
    \label{fig.cyl_Snapshots_b}
    \includegraphics[width=0.33\textwidth, height=0.275\textwidth]
    {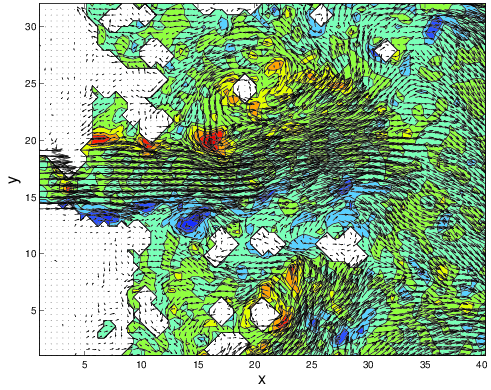} }
  \caption{(a) Geometry of a flow between two cylinders along with the
interrogation window where PIV measurements are taken.  (b,c) Two
    representative snapshots of velocity vectors and vorticity
    contours, resulting from time-resolved PIV measurements,
    illustrate the oscillatory motion of the exiting jet. See text for
    details on the experimental parameters.}
  \label{fig.cyl_Snapshots}
\end{figure}

In this paper, we use a simplified geometric configuration that
nonetheless inherits the main features from more complex settings,
including the presence of multiple distinct frequencies and the
flip-flop oscillations of the exit stream.  In particular, we consider
the flow passing between two cylinders; see
Figure~\ref{fig.cyl_geometry} for a sketch of the geometry. The
PIV interrogation window is positioned about one cylinder radius
downstream of the passage. It measures $40.36\, mm$ in the streamwise
and $32.08\, mm$ in the cross-stream direction. The flow field is
resolved on a $63 \times 79$ measurement grid, and two in-plane
velocity components are recorded. The flow fields are represented in a
fully time-resolved manner with $4\, ms$ between two consecutive PIV
measurements. The inter-cylinder gap is $10.7\, mm$ and the jet
passing between the two cylinders (of diameter $12\, mm$) has a mean
velocity of $0.663\, m/s.$ The resulting Reynolds number, based on the
volume flux velocity ($\dot Q = 18\, m^3/h$) and the cylinder
diameter, is $Re = 3000$.

% [width=0.45\textwidth,height=0.33\textwidth]

Two representative snapshots resulting from a time-resolved flow field
sequence display velocity vectors and contours of the spanwise
vorticity in Figure~\ref{fig.cyl_Snapshots}. The two flow fields are
$0.28\, s$ apart and show a downward
(Figure~\ref{fig.cyl_Snapshots_a}) and upward
(Figure~\ref{fig.cyl_Snapshots_b}) deflection of the jet exiting
between the two cylinders. A time-sequence of flow fields
substantiates the existence of a low-frequency oscillation of the jet,
which forms the basis of the afore-mentioned flip-flop phenomenon. A
streamwise velocity signal has been extracted from the center of the
experimental domain, and a subsequent spectral analysis of this signal
reveals a distinct frequency of $7.81\, \Hz$ which corresponds to a
Strouhal number $St = 0.126$ (based on the jet mean velocity and
cylinder gap).

Figure~\ref{fig.dmd-piv1-amplitude-vs-fr-Re} shows the amplitudes of the DMD modes versus the identified frequencies and corresponding growth/decay rates. These results were obtained using a database with $N = 100$ snapshots. As in the screeching jet example, strongly damped modes are characterized by large amplitudes, and a clear indication about the relative importance of individual DMD modes cannot be inferred from these two plots. Instead, we apply our sparsity-promoting framework to aid in the delineation of modes that contribute significantly to the data sequence and modes that capture only transient effects. By adjusting the user-specified parameter $\gamma,$ we encourage solutions that consist of only a limited number of modes but still optimally represent the original data-sequence.

\begin{figure}
  \centering \subfloat[$N_z \, = \, 21$] {\label{fig.piv1Edmd21}
    \begin{tikzpicture}
      \begin{axis}[ width=0.33\textwidth, height=0.33\textwidth,
        xlabel=$\operatorname{Re} \left( \mu_i \right)$,
        ylabel=$\operatorname{Im} \left( \mu_i \right)$, xmin = -1.1,
        xmax = 1.1, ymin = -1.1, ymax = 1.1 ]
        \addplot[only marks, mark size = 1.5pt, color=black, mark=o, line width = 0.65]
        	table[x=real,y=imag]{data/piv1/piv1Edmd.dat};
        \addplot[only marks, mark size = 1.5pt, color=red, mark=x, line width = 0.65]
        	table[x=real,y=imag]{data/piv1/piv1Edmd21.dat};
	\addplot[no markers, line width = 1.25, color=blue, dashed]
        	table[x=real,y=imag]{data/piv1/circle.dat};
%	\addplot[only marks, mark size = 1.5pt, color=teal, mark=triangle, line width = 0.65]
%        	table[x=real,y=imag]{data/piv1/piv1Elow21.dat};
        \end{axis}
        \end{tikzpicture}
        }
        \hspace*{0.05cm}
        \subfloat[$N_z \, = \, 5$]
        {\label{fig.piv1Edmd5}
        \begin{tikzpicture}
        \begin{axis}[
         width=0.33\textwidth, height=0.33\textwidth,
        	xlabel=$\operatorname{Re} \left( \mu_i \right)$,
	%ylabel=$\operatorname{Im} \left( \mu_i \right)$,
	xmin = 0.9, xmax = 1.002,
	ymin = -0.42, ymax = 0.42
	]
        \addplot[only marks, mark size = 1.5pt, color=black, mark=o, line width = 0.65]
        	table[x=real,y=imag]{data/piv1/piv1Edmd.dat};
        \addplot[only marks, mark size = 1.5pt, color=red, mark=x, line width = 0.65]
        	table[x=real,y=imag]{data/piv1/piv1Edmd5.dat};
	\addplot[no markers, line width = 1.25, color=blue, dashed]
        	table[x=real,y=imag]{data/piv1/circle.dat};
%	\addplot[only marks, mark size = 1.5pt, color=teal, mark=triangle, line width = 0.65]
%        	table[x=real,y=imag]{data/piv1/piv1Elow5.dat};
        \end{axis}
        \end{tikzpicture}
        }
        \hspace*{0.05cm}
        \subfloat[$N_z \, = \, 3$]
        {\label{fig.piv1Edmd3}
        \begin{tikzpicture}
        \begin{axis}[
         width=0.33\textwidth, height=0.33\textwidth,
        	xlabel=$\operatorname{Re} \left( \mu_i \right)$,
	%ylabel=$\operatorname{Im} \left( \mu_i \right)$,
	xmin = 0.9, xmax = 1.002,
	ymin = -0.42, ymax = 0.42
	]
        \addplot[only marks, mark size = 1.5pt, color=black, mark=o, line width = 0.65]
        	table[x=real,y=imag]{data/piv1/piv1Edmd.dat};
        \addplot[only marks, mark size = 1.5pt, color=red, mark=x, line width = 0.65]
        	table[x=real,y=imag]{data/piv1/piv1Edmd3.dat};
	\addplot[no markers, line width = 1.25, color=blue, dashed]
        	table[x=real,y=imag]{data/piv1/circle.dat};
%	\addplot[only marks, mark size = 1.5pt, color=teal, mark=triangle, line width = 0.65]
%        	table[x=real,y=imag]{data/piv1/piv1Elow3.dat};
        \end{axis}
        \end{tikzpicture}
        }
        \caption{Eigenvalues resulting from the standard DMD algorithm
          (circles) along with the subset of $N_z$ eigenvalues
          selected by its sparsity-promoting variant (crosses) for the
          flow through a cylinder bundle. The dashed curves identify
          the unit circle.}
        \label{fig.Edmd-Edmdsp-piv1}
\end{figure}
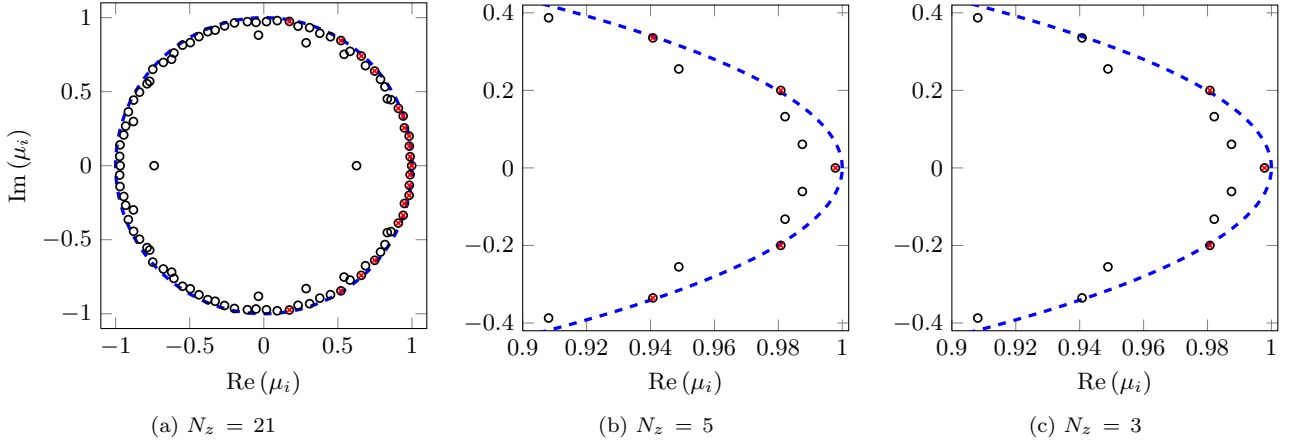

Figure~\ref{fig.Edmd-Edmdsp-piv1} superimposes the eigenvalues
resulting from the standard DMD algorithm (circles) and its
sparsity-promoting variant (crosses). Even though the eigenvalues in
the vicinity of the unit circle are associated with much smaller
amplitudes than the strongly damped eigenvalues (that reside well
within the unit circle), they are selected by our algorithm. As the
sparsity parameter is increased and fewer eigenvalues are selected, a
clustering of eigenvalues at and near the point $(1,0)$ is observed;
see Figure~\ref{fig.piv1Edmd3}.

% Preamble: \pgfplotsset{compat=newest}
\begin{figure}
  \centering \subfloat[] {\label{fig.piv1A_vs_w}
    \begin{tikzpicture}
      \begin{axis}[ width=0.45\textwidth, height=0.35\textwidth,
        xlabel=$\operatorname{Im} \left( \log \, (\mu_i) \right)$,
        ylabel=$| \alpha_i |$, xmin = -3.5, xmax = 3.5, ymin = 0, ymax
        = 115 ]
        \addplot[only marks, mark size = 1.5pt, color=black, mark=o, line width = 0.65]
        	table[x=w,y=Amp] {data/piv1/piv1A_vs_wRe.dat};
        \end{axis}
        \end{tikzpicture}
        }
        \hspace*{0.25cm}
         \subfloat[]
        {\label{fig.piv1A_vs_Re}
        \begin{tikzpicture}
        \begin{axis}[
        width=0.45\textwidth, height=0.35\textwidth,
        	xlabel=$\operatorname{Re} \left( \log \, (\mu_i) \right)$,
	%ylabel=$| \alpha_i |$,
	xmin = -0.5, xmax = 0.01,
	ymin = 0, ymax = 115
	]
        \addplot[only marks, mark size = 1.5pt, color=black, mark=o, line width = 0.65]
        	table[x=real,y=Amp] {data/piv1/piv1A_vs_wRe.dat};
        \end{axis}
        \end{tikzpicture}
        }
        \caption{Dependence of the absolute value of the DMD
          amplitudes $\alpha_i$ on (a) the frequency and (b) the real
          part of the corresponding DMD eigenvalues $\mu_i$ for the
          flow through a cylinder bundle.}
        \label{fig.dmd-piv1-amplitude-vs-fr-Re}
\end{figure}
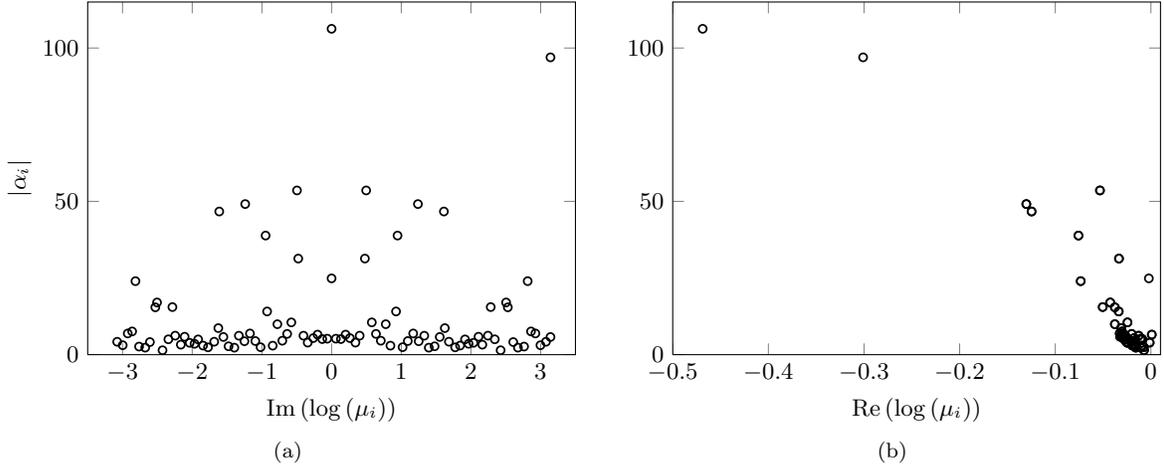

% Preamble: \pgfplotsset{compat=newest}

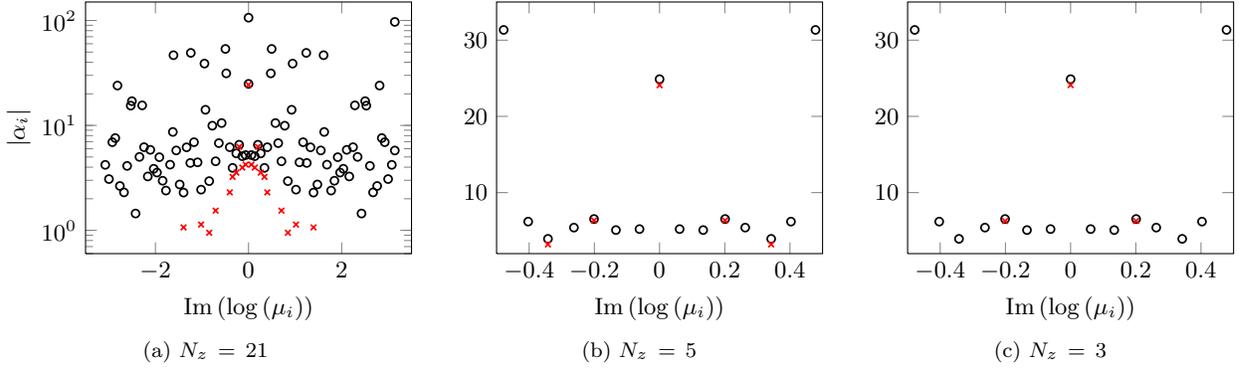
\begin{figure}
  \centering \subfloat[$N_z \, = \, 21$] {\label{fig.piv1Asp_vs_w21}
    \begin{tikzpicture}
      \begin{semilogyaxis}[ width=0.33\textwidth,
        height=0.275\textwidth, xlabel=$\operatorname{Im} \left( \log
          \, (\mu_i) \right)$, ylabel=$| \alpha_i |$, xmin = -3.5,
        xmax = 3.5, ymin = 0.6, ymax = 150 ]
        \addplot[only marks, mark size = 1.5pt, color=black, mark=o, line width = 0.65]
        	table[x=w,y=Amp] {data/piv1/piv1A_vs_wRe.dat};
	\addplot[only marks, mark size = 1.5pt, color=red, mark=x, line width = 0.65]
        	table[x=w,y=Amp] {data/piv1/piv1Asp_vs_wRe21.dat};
        \end{semilogyaxis}
        \end{tikzpicture}
        }
      \hspace*{0.25cm}
      \subfloat[$N_z \, = \, 5$]
        {\label{fig.piv1Asp_vs_w5}
        \begin{tikzpicture}
        \begin{axis}[
        width=0.33\textwidth, height=0.275\textwidth,
        	xlabel=$\operatorname{Im} \left( \log \, (\mu_i) \right)$,
	%ylabel=$| \alpha_i |$,
	xmin = -0.5, xmax = 0.5,
	ymin = 2, ymax = 35
	]
        \addplot[only marks, mark size = 1.5pt, color=black, mark=o, line width = 0.65]
        	table[x=w,y=Amp] {data/piv1/piv1A_vs_wRe.dat};
	\addplot[only marks, mark size = 1.5pt, color=red, mark=x, line width = 0.65]
        	table[x=w,y=Amp] {data/piv1/piv1Asp_vs_wRe5.dat};
        \end{axis}
        \end{tikzpicture}
        }
      \hspace*{0.25cm}
     \subfloat[$N_z \, = \, 3$]
        {\label{fig.piv1Asp_vs_w3}
        \begin{tikzpicture}
        \begin{axis}[
        width=0.33\textwidth, height=0.275\textwidth,
        	xlabel=$\operatorname{Im} \left( \log \, (\mu_i) \right)$,
	%ylabel=$| \alpha_i |$,
	xmin = -0.5, xmax = 0.5,
	ymin = 2, ymax = 35
	]
        \addplot[only marks, mark size = 1.5pt, color=black, mark=o, line width = 0.65]
        	table[x=w,y=Amp] {data/piv1/piv1A_vs_wRe.dat};
	\addplot[only marks, mark size = 1.5pt, color=red, mark=x, line width = 0.65]
        	table[x=w,y=Amp] {data/piv1/piv1Asp_vs_wRe3.dat};
         \end{axis}
        \end{tikzpicture}
        }
        \caption{Dependence of the absolute value of the amplitudes
          $\alpha_i$ on the frequency (imaginary part) of the
          corresponding eigenvalues $\mu_i$ for the flow through a
          cylinder bundle. The results are obtained using the standard
          DMD algorithm (circles) and the sparsity-promoting DMD
          algorithm (crosses) with $N_z$ DMD modes.}
     \label{dmd-vs-dmdsp-piv1-amplitude-vs-fr}
\end{figure}

\begin{figure}
  \begin{center}
    \includegraphics[width=0.45\textwidth,height=0.33\textwidth]{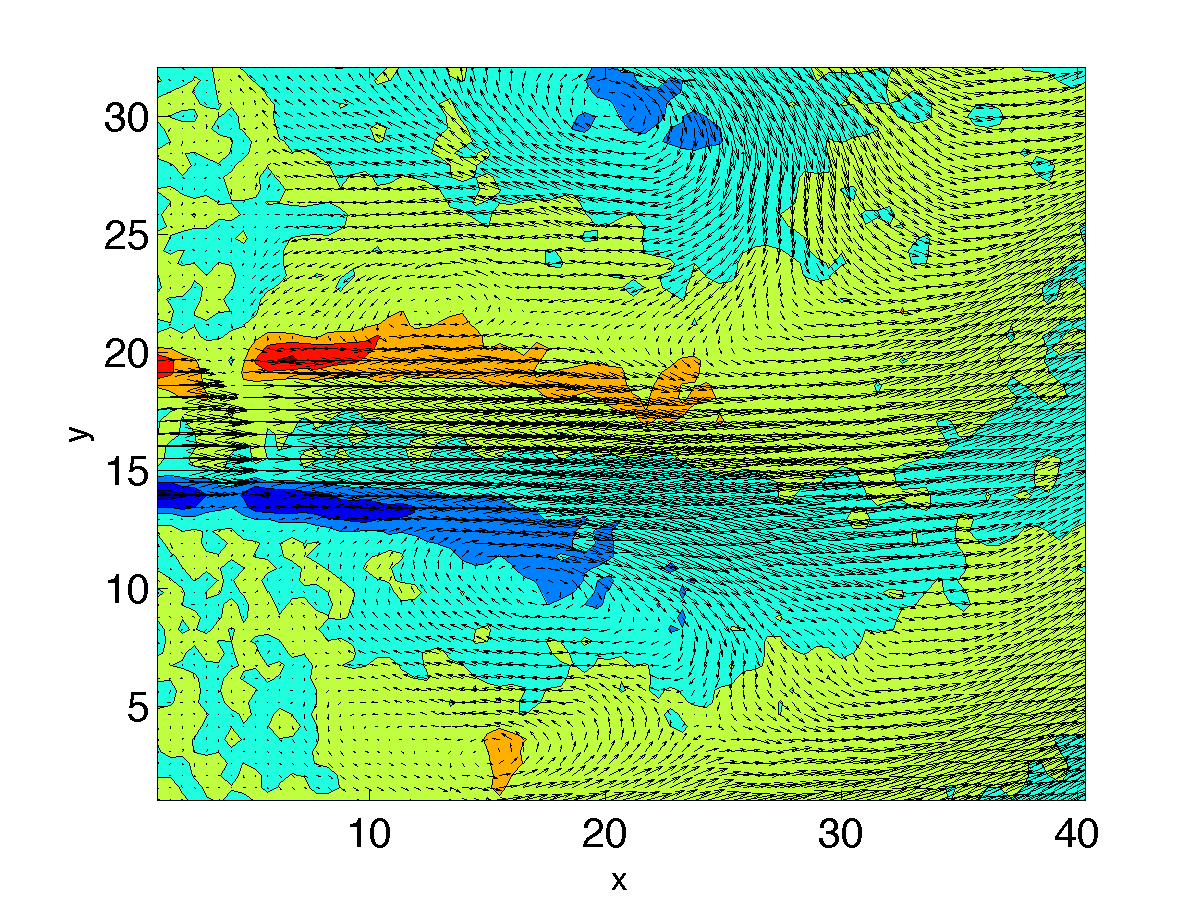}
  \end{center}
  \caption{Snapshot resulting from a sparse three-component ($N_z =
    3$) representation of the full data sequence, visualized by
    velocity vectors and vorticity contours. An animation of this flow
    field is available at \texttt{www.umn.edu/$\sim$mihailo/software/dmdsp/cylinder/}.}
  \label{fig.sparseDMD_Nz3}
\end{figure}

A similar picture emerges from displaying the selected modes in the
frequency-amplitude plane; see
Figure~\ref{dmd-vs-dmdsp-piv1-amplitude-vs-fr}. The sparsity-promoting
DMD algorithm provides a rapid concentration on low-frequency modes,
thereby eliminating structures with substantially larger amplitudes
identified by the standard DMD algorithm. As the limit of only $N_z =
3$ DMD modes is reached, the sparsity-promoting algorithm concentrates
on frequencies that still optimally describe the principal oscillatory
components of the full data set. As evident from
Figure~\ref{fig.piv1Asp_vs_w3}, for $N_z=3$ the last oscillatory
representation is characterized by $\operatorname{Im} \, (\log(\mu_i))
\approx 0.201$ (which corresponds to a frequency of $7.99\, \Hz$ and a
Strouhal number of $St = 0.129$). This frequency is in good agreement
with the frequency identified by the aforementioned point
measurements of the streamwise velocity. The flow field associated
with the three-term ($N_z = 3$) representation of the data sequence is
displayed in Figure~\ref{fig.sparseDMD_Nz3}; it consists of a
deflected jet and strong vortical components off the symmetry axis. An
animation of this flow field, available at \texttt{www.umn.edu/$\sim$mihailo/software/dmdsp/cylinder/}, has been obtained using the identified and selected frequency ($7.99\,
\Hz$) and it reproduces the main features of the full data sequence,
i.e., the lateral swaying of the jet under the influence of the wake
vortices of the two cylinders.

% Preamble: \pgfplotsset{compat=newest}

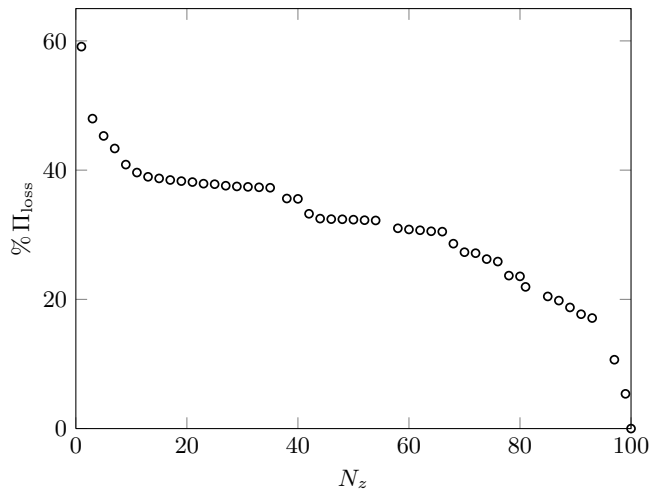
\begin{figure}
  \centering {
    \begin{tikzpicture}
      \begin{axis}[ 
      width=0.5\textwidth, height=0.4\textwidth,
        xlabel=$N_z$, 
        ylabel={$\% \, \Ploss$}, 
        xmin = 0, xmax = 100,
        ymin = 0, ymax = 65
        ]
        \addplot[only marks, mark size = 1.5pt, color=black, mark=o, line width = 0.65]
        	table[x=Nz,y=Ploss] {data/piv1/piv1spPloss_vs_Nz.dat};
%         \addplot[no markers, line width = 1, color=black]
%        	table[x=Nz,y=Ploss] {data/piv1/piv1lowPloss_vs_Nz.dat};
        \end{axis}
        \end{tikzpicture}
        }
        \caption{The optimal performance loss, 
        $ \% \, \Ploss \DefinedAs 100 \, \sqrt{J (\alpha) / J (0)} $, 
        of the sparsity-promoting DMD algorithm for the flow through a
          cylinder bundle as a function of the number of retained modes.}
     \label{fig.piv1-dmdsp-vs-dmdlr}
\end{figure}

The sparse approximation of the full data-sequence focuses on the most relevant structures by removing large-amplitude but transient flow features and it results in progressively larger residuals as the sparsity is enhanced.  This inherent trade-off between a more compact data representation and the quality of approximation compared to the original data sequence is depicted in Figure~\ref{fig.piv1-dmdsp-vs-dmdlr}. In contrast to the channel flow and screeching jet examples, the sparsity-promoting version of the DMD-algorithm displays higher performance degradation. This behavior may perhaps be attributed to the presence of unstructured measurement noise in the experimental dataset for a jet passing between two cylinders.

% Preamble: \pgfplotsset{compat=newest}

	\vspace*{-2ex}
\section{Concluding remarks}
	\vspace*{-1ex}
\label{sec.remarks}

We have introduced an extension of the standard DMD algorithm that addresses the reduction in dimensionality of the full rank decomposition. This reduction is accomplished by a sparsity-promoting procedure which augments the standard least-squares optimization problem by a term that penalizes the $\ell_1$-norm of the vector of unknown amplitudes. A user-specified regularization parameter balances the trade-off between the quality of approximation and the number of retained DMD modes. The sparsity-promoting DMD algorithm thus selects specific dynamic modes (along with their associated temporal frequencies and growth/decay rates) which exhibit the strongest contribution to a representation of the original data sequence over the considered time interval.

The $\ell_1$-regularized least-squares problem can be formulated as a convex optimization problem for which the alternating direction method of multipliers (ADMM) provides an efficient tool for computing the globally optimal solution. ADMM is an iterative method that alternates between minimization of the least-squares residual and sparsity enhancement. We have shown that the least-squares minimization step amounts to solving an unconstrained regularized quadratic program and that sparsity is promoted through the application of a soft-thresholding operator. After a desirable tradeoff between the quality of approximation and the number of DMD modes has been accomplished, we fix the sparsity structure and compute the optimal amplitudes of the retained dynamic modes.

The sparsity-promoting dynamic mode decomposition has been applied to three examples. The linearized plane Poiseuille flow at a supercritical Reynolds number was used to showcase the algorithm on a canonical flow configuration; the sparsity-promoting DMD extracted modal contributions from each of the familiar eigenvalue branches until only the exponentially growing TS wave was retained in the limit of maximal sparsity. For the unstructured large-eddy simulation of a supersonic screeching jet, the sparsity-promoting DMD algorithm successfully and efficiently identified the screech frequency and the associated flow fields; dynamic modes with large amplitudes and substantial decay rates, representing transient flow features, have been eliminated as our emphasis on sparsity increased. Finally, for two-dimensional time-resolved PIV measurements of a jet passing between two cylinders, the prevailing Strouhal number has been identified by the dynamic mode decomposition, and the sparsity-promoting algorithm effectively differentiated the dominant flow structures from less relevant flow phenomena embedded in the full data sequence. 

The developed method builds on recent attempts at optimizing the standard dynamic mode decomposition~\cite{cheturow12,gouwynpeaCDC12,wynpeagangouJFM13} and at providing a framework for the automated detection of a few pertinent modal flow features. While these efforts resulted in complex or nearly intractable non-convex optimization problems, the sparsity-promoting DMD algorithm provides an efficient paradigm for detection and extraction of a limited subset of flow features that optimally approximate the original data sequence. Furthermore, in contrast to the aforementioned attempts, our approach {\em regularizes\/} the least-squares deviation between the matrix of snapshots and the linear combination of DMD modes with a term that penalizes the $\ell_1$-norm of the vector of unknown DMD amplitudes. While the former term depends both on the problem data and on the optimization variable, the latter term is problem-data-independent; its primary purpose is to enhance desirable features in the solution to the resulting regularized optimization problem. We note that regularization may be of essence in the situations where the problem data is corrupted by noise or contains numerical or experimental outliers. In the absence of regularization, the predictive capability of low-dimensional models resulting from corrupted or incomplete snapshots may be significantly diminished. Exploration of different types of regularization penalties, that are well-suited for the problems in fluid mechanics, is an effort worth pursuing. 

The developed algorithm has shown its value on the numerical and experimental snapshot sequences considered in this paper. It is expected that the sparsity-promoting dynamic mode decomposition will become a valuable tool in the quantitative analysis of high-dimensional datasets, in the interpretation of identified dynamic modes, and in the eduction of relevant physical mechanisms.

	\vspace*{-2ex}
\section*{Supplementary material}
	\vspace*{-1ex}

Additional information about the examples considered in this paper, along with {\sc Matlab} source codes and problem data, can be found at \texttt{www.umn.edu/$\sim$mihailo/software/dmdsp/}. 

	\vspace*{-2ex}
\section*{Acknowledgments}
	\vspace*{-1ex}

The authors gratefully acknowledge Prof.\ Parviz Moin for his encouragement to pursue this effort during the 2012 Center for Turbulence Research Summer Program at Stanford University. Supported in part by Stanford University and NASA Ames Research Center, and by the University of Minnesota Initiative for Renewable Energy and the Environment under Early Career Award RC-0014-11. The time-resolved PIV measurements of a jet passing between two cylinders were provided by  Electricit\'e de France (EdF).

\appendix

	\vspace*{-2ex}	
\section{An alternative formulation of~(\ref{eq.alpha_opt})}
	\vspace*{-1ex}
\label{sec.ls-amplitudes}

We show that the objective function $J (\alpha)$ in~(\ref{eq.alpha_opt}),
\beq
    J (\alpha)
    \; = \;
    \|
    G
    \; - \;
    L \, \Dalpha \;\! R
    \|_F^2,
    \label{eq.Jfrob}
\eeq
with
\[
   G
	\; \DefinedAs \;
	\Sigma \, V^*,
	~~
	L
	\; \DefinedAs \;
	Y,
	~~
	\Dalpha
	\; \DefinedAs \;
	\diag \, \{ \alpha \},
	~~
	R
	\; \DefinedAs \;
	\Vand,
\]
can be equivalently represented as~\eqref{eq.Jvec}. The equivalence
between~(\ref{eq.Jfrob}) and~(\ref{eq.Jvec}) can be established
through a sequence of straightforward algebraic manipulations in
conjunction with the repeated use of the following properties of the
matrix trace:
\bi
    \item[{\bf P1:}] Commutativity invariance,
    \[
    \trace
    \left( A \, B \right)
    \; = \,
    \trace
    \left( B \, A \right);
    \]

  \item[{\bf P2:}] A product between a matrix $Q$ and a diagonal
    matrix $\Dalpha \DefinedAs \diag \, \{ \alpha \}$,
    \[
    \trace
    \left( Q \, \Dalpha  \right)
    \; = \,
    \left(
    \diag \, \{ Q \}
    \right)^T
    \!
    \alpha
    \; = \,
    \left(
    \overline{\diag \, \{ Q \}}
    \right)^*
    \!
    \alpha;
    \]

  \item[{\bf P3:}] For vectors $\alpha \in \bbC^n$ and $\beta \in
    \bbC^m$ and matrices $A, B \in \bbC^{m \times n}$,
    \[
    \trace
    \left(
    \Dbeta^*
    \,
    A
    \,
    \Dalpha
    \,
    B^T
    \right)
    \, = \;
    \beta^*
    \!
    \left( A \circ B \right) \alpha.
    \]
\ei
Now, a repeated use of the matrix trace commutativity invariance
property {\bf P1} leads to
\beq
    \ba{rcl}
    J (\alpha)
    & = &
    \|
    G
    \; - \;
    L \, \Dalpha \, R
    \|_F^2
    \\[0.15cm]
    & = &
    \trace
    \left(
    \left(
    G
    \; - \;
    L \, \Dalpha \, R
    \right)^*
    \left(
    G
    \; - \;
    L \, \Dalpha \, R
    \right)
    \right)
    \\[0.15cm]
    & = &
    \trace
    \left(
    \Dalpha^*
    \left(
    L^* L
    \right)
    \Dalpha
    \left(
    R \, R^*
    \right)
    \, - \;
    Q \, \Dalpha
    \; - \;
    Q^*
    \Dalpha^*
    \; + \;
    G^*
    G
    \right),
    \ea
    \non
\eeq
and the equivalence between~(\ref{eq.Jfrob}) and~(\ref{eq.Jvec})
follows from {\bf P2} and {\bf P3}.

	\vspace*{-2ex}
\section{$\alpha$- and $\beta$-minimization steps in ADMM}
	\vspace*{-1ex}
\label{sec.admm}

We exploit the respective structures of the functions $J$ and $g$
in~\eqref{eq.Jell1-vec} and show that the $\alpha$-minimization step
amounts to solving an unconstrained regularized quadratic program and
that the $\beta$-minimization step amounts to an opportune use of the
soft thresholding operator.

\bi
\item {\bf $\alpha$-minimization step:} Completion of squares with
  respect to $\alpha$ in the augmented Lagrangian $\aL$ can be used to
  show that the $\alpha$-minimization step~(\ref{eq.F_update}) is
  equivalent to
  \[
    \ba{rl}
    \underset{\alpha}{\text{minimize}}
    &
    J (\alpha)
    \; + \;
    \dfrac{\rho}{2}
    \,
    \|
    \alpha
    \, - \,
    u^k
    \|_2^2,
    \ea
  \]
where
  \[
    u^k
    \; = \;
    \beta^k
    \; - \,
    \left( 1/\rho \right)
    \lambda^k.
  \]
Now, the definition~(\ref{eq.Jvec}) of $J$ leads to the
unconstrained quadratic programming problem
  \[
    \ba{rl}
    \underset{\alpha}{\text{minimize}}
    &
    \alpha^*
    \!
    \left(
    P \, + \, (\rho/2) \, I
    \right)
    \alpha
    \; - \,
    \left( q \, + \, (\rho/2) \, u^k \right)^* \! \alpha
    \; - \;
    \alpha^* \! \left( q \, + \, (\rho/2) \, u^k \right)
    \; + \;
    s
    \; + \;
    \rho \, \| u^k \|_2^2,
    \ea
  \]
whose solution is determined by
  \[
    \alpha^{k+1}
    \; = \,
    \left(
    P \, + \, (\rho/2) \, I
    \right)^{-1}
    \left( q \, + \, (\rho/2) \, u^k \right).
  \]
\item {\bf $\beta$-minimization step:} The completion of squares with
  respect to $\beta$ in the augmented Lagrangian $\aL$ can be used to
  show that the $\beta$-minimization step~(\ref{eq.G_update}) is
  equivalent to
  \[
    \ba{rl}
    \underset{\beta}{\text{minimize}}
    &
    \gamma \, g (\beta)
    \; + \;
    \dfrac{\rho}{2}
    \,
    \|
    \beta
    \, - \,
    v^k
    \|_2^2,
    \ea
  \]
where
  \[
    v^k
    \; = \;
    \alpha^{k+1}
    \; + \,
    \left( 1/\rho \right)
    \lambda^k.
  \]
It is a standard fact that the solution to this optimization problem
is determined by
  \[
    \beta_i^{k+1}
    \; = \;
    S_{\kappa}
    ( v_i^k )
    ,
    ~~
    \kappa
    \; = \;
    \gamma/\rho,
  \]
where $S_\kappa (\cdot)$ denotes the {\em soft thresholding\/} operator
  \beq
    S_\kappa (v_i^k)
    \; = \;
    \left\{
    \ba{ll}
    v_i^k \,-\, \kappa,
    &
    v_i^k
    \, > \,
    \kappa
    \\[0.15cm]
    0,
    &
    v_i^k
    \, \in \,
    [ -\kappa, \, \kappa ]
    \\[0.15cm]
    v_i^k \,+\, \kappa,
    &
    v_i^k
    \, < \,
    -\kappa.
    \ea
    \right.
    \non
  \eeq

\ei

	\vspace*{-2ex}
\section{An efficient algorithm for solving~(\ref{eq.alpha_opt_sp})}
	\vspace*{-1ex}
\label{sec.ls_sparse}

From Appendix~\ref{sec.ls-amplitudes} it follows that finding the
solution to~(\ref{eq.alpha_opt_sp}) amounts to solving the following
equality-constrained quadratic programming problem

\beq
    \ba{rl}
    \underset{\alpha}{\text{minimize}}
    &
    J (\alpha)
    \; = \;
    \alpha^* P \, \alpha
    \; - \;
    q^* \alpha
    \; - \;
    \alpha^* q
    \; + \;
    s
    \\[0.25cm]
    \text{subject to}
    &
    E^T \, \alpha
    \, = \,
    0.
    \ea
    %%\label{eq.alpha_vec_opt_sp}
    \non
\eeq
It is a standard fact that the variation of the Lagrangian
\beq
    \cL (\alpha,\nu)
    \; = \;
    J (\alpha)
    \; + \:
    \nu^* E^T \alpha
    \; + \,
    \left( E^T \alpha \right)^*
    \!
    \nu,
    \non
\eeq
with respect to $\alpha$ and the vector of Lagrange multipliers $\nu$
can be used to obtain the conditions for optimality of $\cL$,
\beq
   \left[
   \ba{cc}
   {P} & {E}
   \\[0.05cm]
   {E^T} & {0}
   \ea
   \right]
   \left[
   \ba{c}
   {\alpha}
   \\[0.05cm]
   {\nu}
   \ea
   \right]
   \; = \;
   \left[
   \ba{c}
   {q}
   \\[0.1cm]
   {0}
   \ea
   \right].
   \non
\eeq
The optimal sparse vector of amplitudes, $\alpha_{\mathrm{sp}}$, is
then determined by
\beq
   \alpha_{\mathrm{sp}}
   \; = \;
   \left[
   \ba{cc}
   {I} & {0}
   \ea
   \right]
   \left[
   \ba{cc}
   {P} & {E}
   \\[0.05cm]
   {E^T} & {0}
   \ea
   \right]^{-1}
   \left[
   \ba{c}
   {q}
   \\[0.1cm]
   {0}
   \ea
   \right].
   \non
\eeq

% ==================== THE BIBLIOGRAPHY ====================


\begin{thebibliography}{10}
\providecommand{\url}[1]{#1}
\csname url@samestyle\endcsname
\providecommand{\newblock}{\relax}
\providecommand{\bibinfo}[2]{#2}
\providecommand{\BIBentrySTDinterwordspacing}{\spaceskip=0pt\relax}
\providecommand{\BIBentryALTinterwordstretchfactor}{4}
\providecommand{\BIBentryALTinterwordspacing}{\spaceskip=\fontdimen2\font plus
\BIBentryALTinterwordstretchfactor\fontdimen3\font minus
  \fontdimen4\font\relax}
\providecommand{\BIBforeignlanguage}[2]{{%
\expandafter\ifx\csname l@#1\endcsname\relax
\typeout{** WARNING: IEEEtran.bst: No hyphenation pattern has been}%
\typeout{** loaded for the language `#1'. Using the pattern for}%
\typeout{** the default language instead.}%
\else
\language=\csname l@#1\endcsname
\fi
#2}}
\providecommand{\BIBdecl}{\relax}
\BIBdecl

\bibitem{lumley}
J.~Lumley, \emph{Stochastic Tools in Turbulence}.\hskip 1em plus 0.5em minus
  0.4em\relax Dover Publications, 2007.

\bibitem{sirovich}
L.~Sirovich, ``Turbulence and the dynamics of coherent structures. {P}art {I}:
  {C}oherent structures,'' \emph{Quart. Appl. Math.}, vol. 45(3), pp. 561--571,
  1987.

\bibitem{sipp}
D.~Sipp, O.~Marquet, P.~Meliga, and A.~Barbagallo, ``Dynamics and control of
  global instabilities in open flows: a linearized approach,'' \emph{Appl.
  Mech. Rev.}, vol.~63, p. 030801, 2010.

\bibitem{moore}
B.~Moore, ``Principal component analysis in linear systems: controllability,
  observability and model reduction,'' \emph{IEEE Trans. Automat. Control},
  vol. AC-26(1), 1981.

\bibitem{rowley}
C.~Rowley, ``Model reduction for fluids using balanced proper orthogonal
  decomposition,'' \emph{Int. J. Bifurcation Chaos}, vol.~15, pp. 997--1013,
  2005.

\bibitem{Mezic05}
I.~Mezi{\'c}, ``Spectral properties of dynamical systems, model reduction and
  decompositions,'' \emph{Nonlinear Dynamics}, vol.~41, no.~1, pp. 309--325,
  2005.

\bibitem{rowley2}
C.~Rowley, I.~Mezic, S.~Bagheri, P.~Schlatter, and D.~Henningson, ``Spectral
  analysis of nonlinear flows,'' \emph{J. Fluid Mech.}, vol. 641, pp. 115--127,
  2009.

\bibitem{Mezic13}
I.~Mezi\'c, ``Analysis of fluid flows via spectral properties of {K}oopman
  operator,'' \emph{Ann. Rev. Fluid Mech.}, vol.~45, no.~1, pp. 357--378, 2013.

\bibitem{schJFM10dmd}
P.~J. Schmid, ``Dynamic mode decomposition of numerical and experimental
  data,'' \emph{J. Fluid Mech.}, vol. 656, pp. 5--28, 2010.

\bibitem{tretrereddri93}
L.~N. Trefethen, A.~E. Trefethen, S.~C. Reddy, and T.~A. Driscoll,
  ``Hydrodynamic stability without eigenvalues,'' \emph{Science}, vol. 261, pp.
  578--584, 1993.

\bibitem{jovbamJFM05}
M.~R. Jovanovi\'c and B.~Bamieh, ``Componentwise energy amplification in
  channel flows,'' \emph{J. Fluid Mech.}, vol. 534, pp. 145--183, July 2005.

\bibitem{Schmid2007}
P.~J. Schmid, ``Nonmodal stability theory,'' \emph{Annu. Rev. Fluid Mech.},
  vol.~39, pp. 129--162, 2007.

\bibitem{bag13}
S.~Bagheri, ``Koopman-mode decomposition of the cylinder wake,'' \emph{J. Fluid
  Mech.}, vol. 726, pp. 596--623, 2013.

\bibitem{cheturow12}
K.~K. Chen, J.~H. Tu, and C.~W. Rowley, ``Variants of dynamic mode
  decomposition: boundary condition, {K}oopman, and {F}ourier analyses,''
  \emph{J. Nonlinear Sci.}, vol.~22, no.~6, pp. 887--915, 2012.

\bibitem{gouwynpeaCDC12}
P.~J. Goulart, A.~Wynn, and D.~Pearson, ``Optimal mode decomposition for high
  dimensional systems,'' in \emph{51st IEEE Conference on Decision and
  Control}, 2012, pp. 4965--4970.

\bibitem{wynpeagangouJFM13}
A.~Wynn, D.~Pearson, B.~Ganapathisubramani, and P.~J. Goulart, ``Optimal mode
  decomposition for unsteady flows,'' \emph{J. Fluid Mech.}, 2013, to appear.

\bibitem{boyd}
S.~Boyd and L.~Vandenberghe, \emph{Convex optimization}.\hskip 1em plus 0.5em
  minus 0.4em\relax Cambridge University Press, 2004.

\bibitem{canromtao06}
E.~J. Cand\`{e}s, J.~Romberg, and T.~Tao, ``Robust uncertainty principles:
  Exact signal reconstruction from highly incomplete frequency information,''
  \emph{IEEE Trans. Inf. Theory}, vol.~52, no.~2, pp. 489--509, 2006.

\bibitem{don06}
D.~L. Donoho, ``Compressed sensing,'' \emph{IEEE Trans. Inf. Theory}, vol.~52,
  no.~4, pp. 1289--1306, 2006.

\bibitem{cantao06}
E.~J. Cand\`{e}s and T.~Tao, ``Near optimal signal recovery from random
  projections: Universal encoding strategies?'' \emph{IEEE Trans. Inf. Theory},
  vol.~52, no.~12, pp. 5406--5425, 2006.

\bibitem{canwakboy08}
E.~J. Cand\`{e}s, M.~B. Wakin, and S.~P. Boyd, ``Enhancing sparsity by
  reweighted $\ell_1$ minimization,'' \emph{J. Fourier Anal. Appl}, vol.~14,
  pp. 877--905, 2008.

\bibitem{hastibfri09}
T.~Hastie, R.~Tibshirani, and J.~Friedman, \emph{The elements of statistical
  learning}.\hskip 1em plus 0.5em minus 0.4em\relax Springer, 2009.

\bibitem{boyparchupeleck11}
S.~Boyd, N.~Parikh, E.~Chu, B.~Peleato, and J.~Eckstein, ``Distributed
  optimization and statistical learning via the alternating direction method of
  multipliers,'' \emph{Foundations and Trends in Machine Learning}, vol.~3,
  no.~1, pp. 1--124, 2011.

\bibitem{linfarjovTAC13admm}
F.~Lin, M.~Fardad, and M.~R. Jovanovi\'c, ``Design of optimal sparse feedback
  gains via the alternating direction method of multipliers,'' \emph{IEEE
  Trans. Automat. Control}, vol.~58, no.~9, pp. 2426--2431, 2013.

\bibitem{cvx}
M.~Grant and S.~Boyd, ``{CVX}: Matlab software for disciplined convex
  programming, version 2.0 beta,'' \url{http://cvxr.com/cvx}, 2012.

\bibitem{schhen01}
P.~J. Schmid and D.~S. Henningson, \emph{Stability and Transition in Shear
  Flows}.\hskip 1em plus 0.5em minus 0.4em\relax Springer-Verlag, 2001.

\bibitem{weired00}
J.~A.~C. Weideman and S.~C. Reddy, ``A {MATLAB} differentiation matrix suite,''
  \emph{{ACM} Transactions on Mathematical Software}, vol.~26, no.~4, pp.
  465--519, December 2000.

\bibitem{tam95}
C.~K.~W. Tam, ``Supersonic jet noise,'' \emph{Annu. Rev. Fluid Mech.}, vol.~27,
  pp. 17--43, 1995.

\bibitem{frate11}
F.~C. Frate and J.~E. Bridges, ``Extensible rectangular nozzle model system,''
  AIAA Paper 2011-975, 2011.

\bibitem{nichols11}
J.~W. Nichols, F.~E. Ham, and S.~K. Lele, ``High-fidelity large-eddy simulation
  for supersonic rectangular jet noise prediction,'' AIAA Paper 2011-2919,
  2011.

\bibitem{constantine11}
P.~G. Constantine and D.~F. Gleich, ``Tall and skinny {QR} factorizations in
  {M}ap{R}educe architectures,'' in \emph{Proceedings of the 2nd international
  workshop on MapReduce and its applications}, 2011, pp. 43--50.

\bibitem{kakac:97}
S.~Kakac and H.~Liu, \emph{Heat Exchangers: Selection, Rating and Thermal
  Design}.\hskip 1em plus 0.5em minus 0.4em\relax CRC Press, 1997.

\bibitem{rollet-miet:1999}
P.~Rollet-Miet, D.~Laurence, and J.~Ferziger, ``{LES} and {RANS} of turbulent
  flow in tube bundles,'' \emph{Int. J. Heat and Fluid Flow}, vol.~20, pp.
  241--254, 1999.

\bibitem{sumner:2000}
D.~Sumner, S.~Price, and M.~Paidoussis, ``Flow-pattern identification for two
  staggered circular cylinders in cross-flow,'' \emph{J. Fluid Mech.}, vol.
  411, pp. 263--303, 2000.

\bibitem{benhamadouche:2003}
S.~Benhamadouche and D.~Laurence, ``{LES}, coarse {LES}, and transient {RANS}
  comparisons on the flow across a tube bundle,'' \emph{Int. J. Heat and Fluid
  Flow}, vol.~24, pp. 470--479, 2003.

\bibitem{moulinec:2004}
C.~Moulinec, M.~Pourqui\'e, B.~Boersma, T.~Buchal, and F.~Nieuwstadt, ``Direct
  numerical simulation on a {C}artesian mesh of the flow through a tube
  bundle,'' \emph{Int. J. Comp. Fluid Dyn.}, vol.~18, pp. 1--14, 2004.

\bibitem{liang:2007}
C.~Liang and G.~Papadakis, ``Large eddy simulation of cross-flow through a
  staggered tube bundle at subcritical {R}eynolds number,'' \emph{J. Fluids
  Struct.}, vol.~23, pp. 1215--1230, 2007.

\bibitem{simonin}
O.~Simonin and M.~Barcouda, ``Measurements and prediction of turbulent flow
  entering a staggered tube bundle,'' in \emph{Proceedings of the 4th
  International Symposium on Applications of Laser Anemometry to Fluid
  Mechanics, Lisbon}, 1988, p. 5.23.

\bibitem{hassan:1999}
Y.~Hassan and H.~Barsamian, ``Turbulence simulation in tube bundle geometries
  using the dynamic subgrid-scale model,'' \emph{Nuclear Techn. J.}, vol. 128,
  pp. 58--74, 1999.

\end{thebibliography}
\end{document}